\documentclass[amsmath,amssymb]{revtex4}
\usepackage{graphicx}
\usepackage{amscd,amsmath,amsthm,amsfonts,amssymb}
\usepackage{ytableau}
\usepackage{extarrows}
\def\[{\begin{equation}}
\def\]{\end{equation}}

\begin{document}
\title{Rogue curves in the Davey-Stewartson I equation}
\author{
Bo Yang$^{1}$, Jianke Yang$^{2}$}
\address{$^{1}$ School of Mathematics and Statistics, Ningbo University, Ningbo 315211, China\\
$^{2}$ Department of Mathematics and Statistics, University of Vermont, Burlington, VT 05401, U.S.A}
\begin{abstract}
We report new rogue wave patterns whose wave crests form closed or open curves in the spatial plane, which we call rogue curves, in the Davey-Stewartson I equation. These rogue curves come in various striking shapes, such as rings, double rings, and many others. They emerge from a uniform background (possibly with a few lumps on it), reach high amplitude in such striking shapes, and then disappear into the same background again. We reveal that these rogue curves would arise when an internal parameter in bilinear expressions of the rogue waves is real and large. Analytically, we show that these rogue curves are predicted by root curves of certain types of double-real-variable polynomials. We compare analytical predictions of rogue curves to true solutions and demonstrate good agreement between them.
\end{abstract}
\maketitle

\section{Introduction}
Evolution of a two-dimensional wave packet on water of finite depth is governed by the Benney-Roskes-Davey-Stewartson equation \cite{Benney_Roskes,Davey_Stewartson,Ablowitz_book}. In the shallow water limit, this equation is integrable (see Ref. \cite{Ablowitz_book} and the references therein). This integrable equation is sometimes just called the Davey-Stewartson (DS) equation in the literature. The DS equation is divided into two types, DSI and DSII, which correspond to the strong surface tension and weak surface tension, respectively \cite{Ablowitz_book}.

Rogue waves are large spontaneous and unexpected wave excitations. They mysteriously appear from a certain background (uniform or not), rise to high amplitudes, and then retreat back to the same background again. Due to their mysterious nature, a lot of experimental work has been done on rogue waves in diverse physical systems, such as water waves \cite{PeliBook,Tank1,Tank2, Tank3, Tank4}, optical waves \cite{Solli_Nature,Wabnitz_book,Fiber1,Fiber1b}, plasma \cite{Plasma}, Bose-Einstein condensates \cite{RogueBEC}, acoustics \cite{Acoustics}, etc.

Theoretically, many of such rogue waves can be described by rational solutions of certain integrable equations, such as the nonlinear Schr\"odinger (NLS) equation, the Manakov system, and so on. Those rational solutions constitute theoretical rogue waves, and their expressions have been derived in a wide variety of integrable systems, including the NLS equation \cite{Peregrine,AAS2009,DGKM2010,KAAN2011,GLML2012,OhtaJY2012,PeliNLS}, the Manakov system \cite{BDCW2012,ManakovDark,LingGuoZhaoCNLS2014,Chen_Shihua2015,ZhaoGuoLingCNLS2016}, and many others.

Rogue waves in DSI have been studied in \cite{OhtaYangDSI}. It was found that the fundamental rogue waves have wave crests in straight lines in the spatial plane. Certain higher-order rogue waves describe the nonlinear interaction of multiple fundamental rogue waves, while some other higher-order rogue waves exhibit parabola-shaped wave crests in the spatial plane.

In this paper, we report new rogue patterns whose wave crests form closed or open curves in the spatial plane, which we call rogue curves. These rogue curves come in various interesting shapes, such as rings, double rings, and many others. They arise from a uniform background (possibly with a few lumps on it), reach high amplitude in such interesting shapes, and then disappear into the same background again. We will present these rogue curves in the context of the DSI equation. Importantly, we find that such rogue curves would appear when an internal parameter in bilinear expressions of the rogue waves are real and large. Performing large-parameter asymptotic analysis, we show that these rogue curves can be predicted by root curves of certain types of double-real-variable polynomials. We compare our analytical predictions of rogue curves to true solutions and demonstrate good agreement between them.

\section{Preliminaries}
The Davey-Stewartson-I (DSI) equation is
\begin{equation} \label{DS}
\begin{array}{l}
\textrm{i}A_t=A_{xx} + A_{yy}+ (\epsilon|A|^2-2 Q)A,
\\[5pt]
Q_{xx}-Q_{yy}=\epsilon (|A|^2)_{xx},
\end{array}
\end{equation}
where $\epsilon=\pm 1 $ is the sign of nonlinearity. Rogue wave solutions in this equation have been presented in \cite{OhtaYangDSI}. But those solutions involve differential operators and are not explicit. Explicit expressions of those rogue waves and their proof will be presented in the appendix.

The rogue waves in the appendix (and in \cite{OhtaYangDSI}) contain various types of solutions, such as multi-rogue waves and higher-order rogue waves, depending on whether the spectral parameters in them are the same or different. In addition, those solutions contain many free internal parameters. In this article, we consider the higher-order rogue waves where all the spectral parameters are the same, and those internal free parameters are under certain restrictions. Explicit expressions of these higher-order rogue waves are much simpler. To present these solutions, we first introduce elementary Schur polynomials $S_n(\mbox{\boldmath $x$})$ with $ \emph{\textbf{x}}=\left( x_{1}, x_{2}, \ldots \right)$, which are defined by the generating function
\begin{equation}\label{Elemgenefunc}
\sum_{n=0}^{\infty}S_n(\mbox{\boldmath $x$}) \epsilon^n
=\exp\left(\sum_{k=1}^{\infty}x_k \epsilon^k\right).
\end{equation}
We also define $S_n=0$ if $n<0$. Then, these higher-order rogue waves are given by the following lemma.

\begin{quote}
\textbf{Lemma 1} The Davey-Stewartson I eqaution (\ref{DS}) admits higher-order rogue wave solutions
\begin{eqnarray}
  && A_{\Lambda}(x,y,t)= \sqrt{2}\frac{g}{f}, \label{DSRWsolu1} \\
  && Q_{\Lambda}(x,y,t)= 1-2 \epsilon \left( \log f \right)_{xx}, \label{DSRWsolu2}
\end{eqnarray}
where $\Lambda=(n_1, n_2, \dots, n_N)$ is an order-index vector, $N$ is the length of $\Lambda$, each $n_i$ is a nonnegative integer, $n_1<n_2<\cdots <n_N$,
\[ \label{diffopesolufN}
f=\tau_{0}, \quad g=\tau_{1},
\]
\[ \label{deftaunk}
\tau_{k}=
\det_{
\begin{subarray}{l}
1\leq i, j \leq N
\end{subarray}
}
\left(
\begin{array}{c}
m_{i,j}^{(k)}
\end{array}
\right),
\]
the matrix elements $m_{i,j}^{(k)}$ of $\tau_{k}$ are defined by
\[ \label{Schmatrimnij}
m_{i,j}^{(k)}=\sum_{\nu=0}^{\min(n_{i}, n_{j})}\frac{1}{4^\nu} \hspace{0.06cm} S_{n_{i}-\nu}[\textbf{\emph{x}}^{+}(k) +\nu \textbf{\emph{s}} ] \hspace{0.06cm} S_{n_{j}-\nu}[\textbf{\emph{x}}^{-}(k) + \nu \textbf{\emph{s}}],
\]
vectors $\textbf{\emph{x}}^{\pm}(k)=\left( x_{1}^{\pm}, x_{2}^{\pm},\cdots \right)$ are
\begin{eqnarray}
&&x_{r}^{+}(k)= \frac{(-1)^r}{r!p} x_{-1} + \frac{(-2)^r}{r!p^2} x_{-2} + \frac{1}{r!} p x_1 + \frac{2^r}{r!} p^2 x_2  + k\delta_{r, 1}+a_r,    \label{xrijplus}\\
&&x_{r}^{-}(k)= \frac{(-1)^r}{r!p} x_{-1} + \frac{(-2)^r}{r!p^2} x_{2} + \frac{1}{r!} p x_1 + \frac{2^r}{r!} p^2 x_{-2}  - k \delta_{r,1}+a_r^*,
\end{eqnarray}
\begin{equation} \label{tranDSI}
\begin{array}{ll}
x_1=\frac{1}{2}(x+y), & x_{-1}=\frac{1}{2}\epsilon (x-y), \\ [5pt]
x_{2}=-\frac{1}{2}\textrm{i}t, & x_{-2}=\frac{1}{2}\textrm{i}t,
\end{array}
\end{equation}
$p$ is a real constant, $\delta_{r, 1}$ is the Kronecker delta function which is equal to 1 when $r=1$ and 0 otherwise, $\textbf{\emph{s}}=(0, s_2, 0, s_4, \cdots)$ are coefficients from the expansion
\[
\ln \left[\frac{2}{\kappa} \tanh \frac{\kappa}{2}\right] = \sum_{r=1}^{\infty}s_{r} \kappa^r, \label{schurcoeffsr}
\]
and $a_{1}, a_2, \dots, a_{n_N}$ are free complex constants.
\end{quote}

This lemma will be derived at the end of the appendix.

The free internal parameter $a_1$ can be absorbed into $(x, t)$ or $(y, t)$ through a coordinate shift. In addition, under the variable transformation of $Q \rightarrow Q + \epsilon |A|^2,\ x \leftrightarrow y$, and $\epsilon \rightarrow - \epsilon $, the DSI equation (\ref{DS}) is invariant. Thus, we will set
\[
\epsilon=1, \quad a_1=0,
\]
in this article without loss of generality. In addition, we denote $\textbf{\emph{a}}=(0, a_2, \dots, a_{n_N})$.

\section{Rogue curves}

To demonstrate rogue curves in DSI, we first show two examples. In the first example, we choose
\[ \label{paraDSI1}
p=1, \quad \Lambda=(1,4), \quad \textbf{\emph{a}}=(0,0,0, 5000).
\]
The corresponding solution $|A|$ from Lemma 1 at four time values of $t=-3, -1, 0$ and 3 is shown in Fig.~1. It is seen that a rogue wave in the shape of two separate curves symmetric with respect to the $x$-axis arise from the uniform background in the $(x, y)$ plane. These rogue curves reach peak amplitude of $3\sqrt{2}$ at $t=0$, and then retreat to the same uniform background again. The shapes of these rogue curves are not parabolas but more complex, and their appearance is mysterious.

\begin{figure}[htb]
\begin{center}
\includegraphics[scale=0.5, bb=550 0 300 300]{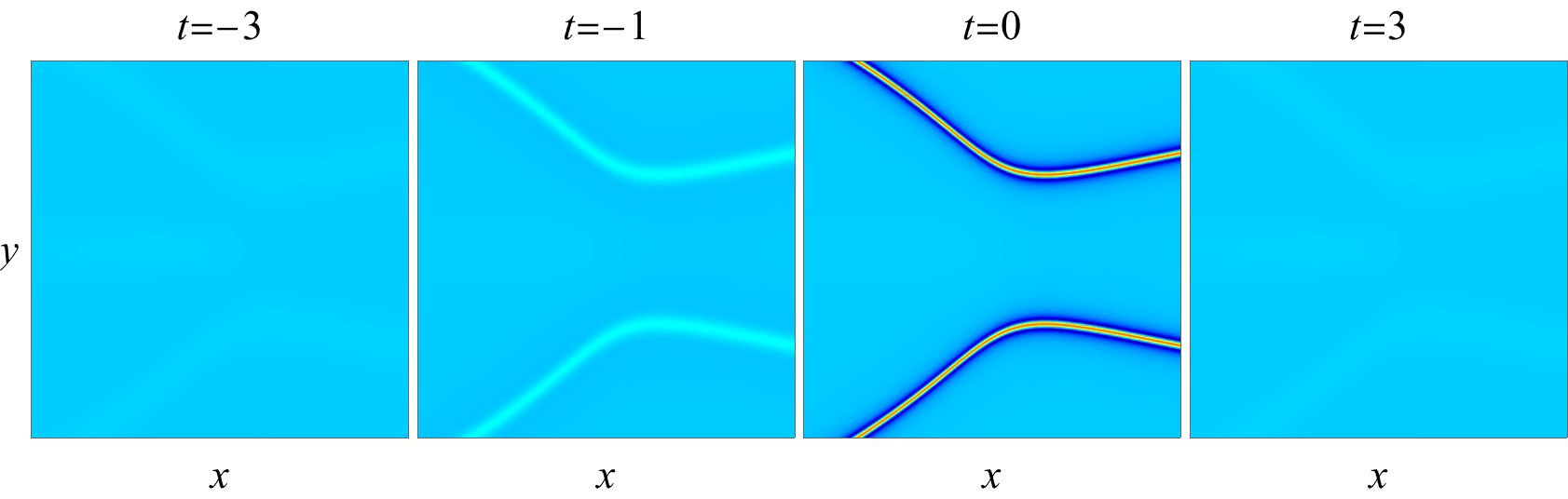}
\caption{A rogue curve ($|A|$) in the DSI equation at four time values of $t=-3, -1, 0$ and 3 for parameter choices in Eq.~(\ref{paraDSI1}). In all panels, $-500\le x \le 500$, and $-30\le y\le 30$. } \label{f:roguecurve}
\end{center}
\end{figure}

An even more interesting example comes when we choose
\[ \label{paraDSI2}
p=1, \quad \Lambda=(2,3), \quad  \textbf{\emph{a}}=(0,0,2000),
\]
and the corresponding solution $|A|$ from Lemma 1 at four time values of $t=-4, -2, 0$ and 4 is shown in Fig.~2. It is seen that at large times ($t=\pm 4$), the solution contains two lumps on the uniform background. But at the intermediate time of $t=-2$, a rogue wave whose crests form a closed curve in the $(x, y)$ plane starts to appear between the two lumps (we call this rogue closed curve a rogue ring). This rogue ring reaches peak amplitude of $3\sqrt{2}$ at $t=0$, after which it starts to disappear and becomes invisible when $t=4$. The appearance of this rogue ring is more mysterious.

\begin{figure}[htb]
\begin{center}
\includegraphics[scale=0.5, bb=550 0 300 300]{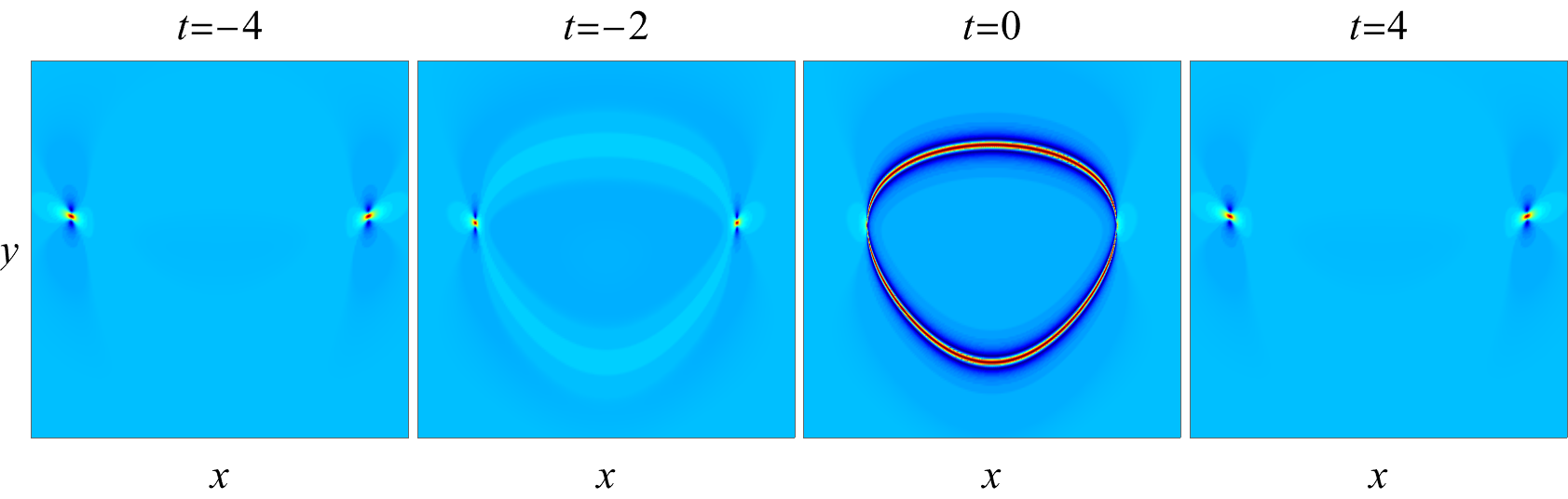}
\caption{A rogue ring ($|A|$) in the DSI equation at four time values of $t=-4, -2, 0$ and 4 for parameter choices in Eq.~(\ref{paraDSI2}). In all panels, $-500\le x \le 500$, and $-10\le y\le 40$. } \label{f:roguering}
\end{center}
\end{figure}

How can we understand these rogue curves? In particular, how can we analytically predict the shapes and locations of these rogue curves? This will be done in the next section.

\section{Asymptotic prediction of rogue curves}

It turns out that these rogue curves in the previous section can be predicted by root curves of certain types of double-real-variable polynomials. So we introduce such polynomials and their root curves first.

\subsection{Special double-real-variable polynomials and their root curves}
We introduce a class of special polynomials in two real variables $(z_1, z_2)$, which can be written as a determinant
\begin{eqnarray} \label{DoubleRealPolydef}
&& \mathcal{P}_{\Lambda}^{[m]}(z_1, z_2) = \left| \begin{array}{cccc}
         \mathcal{S}^{[m]}_{n_1}(z_1, z_2) & \mathcal{S}^{[m]}_{n_1-1}(z_1, z_2) & \cdots &  \mathcal{S}^{[m]}_{n_1-N+1}(z_1, z_2) \\
         \mathcal{S}^{[m]}_{n_2}(z_1, z_2) & \mathcal{S}^{[m]}_{n_2-1}(z_1, z_2) & \cdots &  \mathcal{S}^{[m]}_{n_2-N+1}(z_1, z_2) \\
        \vdots& \vdots & \vdots & \vdots \\
         \mathcal{S}^{[m]}_{n_N}(z_1, z_2) & \mathcal{S}^{[m]}_{n_N-1}(z_1, z_2) & \cdots &  \mathcal{S}^{[m]}_{n_N-N+1}(z_1, z_2)
       \end{array}
 \right|,
\end{eqnarray}
where $\mathcal{S}^{[m]}_{k}(z_1, z_2)$ are Schur polynomials in two variables defined by
\begin{equation}\label{AMthetak}
\sum_{k=0}^{\infty} \mathcal{S}^{[m]}_k(z_1, z_2) \epsilon^k =\exp\left( z_2 \epsilon + z_1 \epsilon^{2} +  \epsilon^{m}  \right), \ \ \ m\geq3,
\end{equation}
$\Lambda=(n_1, n_2, \dots, n_N)$ is an order-index vector, and $\mathcal{S}^{[m]}_{k}(z_1, z_2)\equiv 0$ if $k<0$. This determinant is a Wronskian (in $z_2$) since we can see from Eq.~(\ref{AMthetak}) that
\[ \label{Smdz2}
\frac{\partial}{\partial z_2}\mathcal{S}^{[m]}_{k}(z_1, z_2)=\mathcal{S}^{[m]}_{k-1}(z_1, z_2).
\]
A few such polynomials are given below by choosing specific $m$ and $\Lambda$ values,
\begin{eqnarray}
&& m=4, \Lambda=(1,4), \ \ \ \mathcal{P}^{[m]}_{\Lambda}(z_1, z_2)=\left( z_2^4 +4z_1z_2^2 -4z_1^2 -8\right)/8, \label{para1}\\
&& m=3, \Lambda=(2,3),\ \ \ \mathcal{P}^{[m]}_{\Lambda}(z_1, z_2)=\left(z_2^4 + 12 z_1^2 - 12z_2\right)/12, \\
&& m=4, \Lambda=(2,4),\ \ \ \mathcal{P}^{[m]}_{\Lambda}(z_1, z_2)=z_2 \left(z_2^4+4z_1z_2^2+ 12 z_1^2-24\right)/24,  \\
&& m=5, \Lambda=(4,5),\ \ \ \mathcal{P}^{[m]}_{\Lambda}(z_1, z_2)=\left(z_2^8+ 16 z_1 z_2^6+ 120 z_1^2 z_2^4 + 720 z_1^4 - 480 z_2^3 - 2880z_1 z_2\right)/2880. \label{para4}
\end{eqnarray}

\begin{figure}[htb]
\begin{center}
\includegraphics[scale=0.6, bb=300 000 500 200]{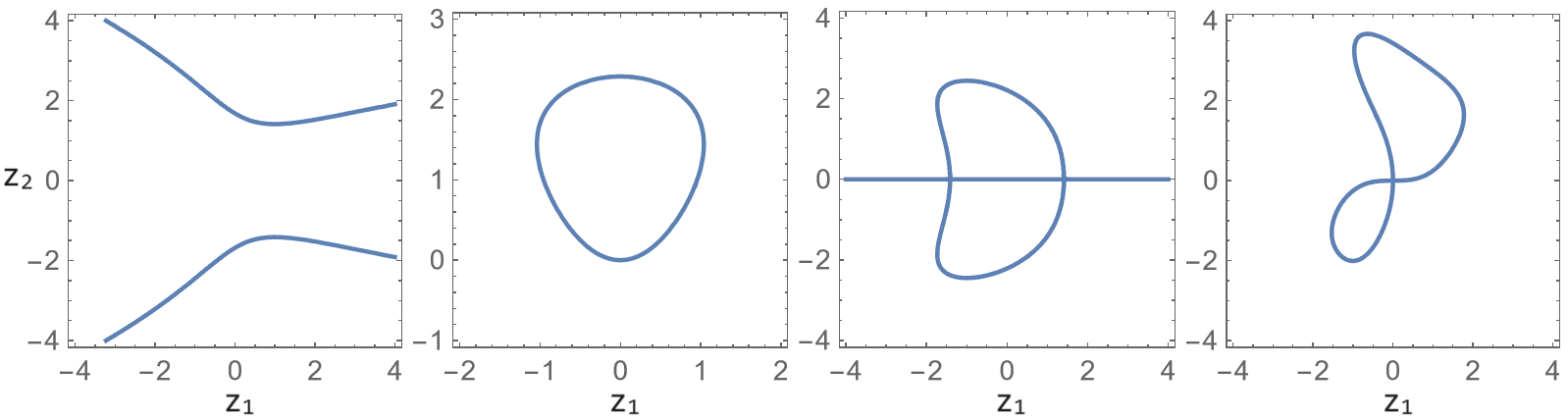}
\caption{Root curves $z_2=\mathcal{R}_{\Lambda, m}(z_1)$ of the double-real-variable polynomial $\mathcal{P}_{\Lambda}^{[m]}(z_1, z_2)$ in the $(z_1, z_2)$ plane for parameter choices in Eqs.~(\ref{para1})-(\ref{para4}), from left to right, respectively. } \label{f:roots}
\end{center}
\end{figure}

By setting
\[ \label{Pmz1z20}
\mathcal{P}^{[m]}_{\Lambda}(z_1, z_2)=0
\]
for real values of $(z_1, z_2)$, we get root curves of this equation in the $(z_1, z_2)$ plane. Let us denote these root curve solutions as
\[ \label{Pmz1z20b}
z_2=\mathcal{R}_{\Lambda, m}(z_1).
\]
For the above four examples of $\mathcal{P}^{[m]}_{\Lambda}(z_1, z_2)$, their root curves are displayed in Fig.~\ref{f:roots}. As one can see, these root curves may be an open curve, as in the first example, or a closed curve, as in the second and fourth examples, or a mixture of open and closed curves, as in the third example. For closed curves, they can be a single loop as in the second example, or a connected double loop as in the fourth example. Other varieties of these curves are also possible for other examples of $\mathcal{P}^{[m]}_{\Lambda}(z_1, z_2)$, such as disconnected double loops and so on.

On a root curve, there may exist some special points where
\[ \label{dPmz1z20}
\frac{\partial\mathcal{P}^{[m]}_{\Lambda}(z_1, z_2)}{\partial z_2}=0.
\]
Such special points will be important to us, and we will call them exceptional points of the root curve. These exceptional points satisfy both Eqs.~(\ref{Pmz1z20}) and (\ref{dPmz1z20}). To easily see where these exceptional points are located on a root curve, it is helpful to consider the dynamical system
\[ \label{dz2dt}
\frac{dz_2}{dt}=\mathcal{P}^{[m]}_{\Lambda}(z_1, z_2),
\]
where $z_2$ is treated as a real function of time $t$, and $z_1$ is treated as a real parameter. For this dynamical system, the root curve (\ref{Pmz1z20b}) gives its bifurcation diagram, while Eq.~(\ref{dPmz1z20}) is the bifurcation condition on this diagram. From this point of view, it is then clear that the exceptional points of the root curve are the bifurcation points of this root curve (when this root curve is viewed as a bifurcation diagram). This realization then makes it very easy to identify exceptional points of the root curve. For example, on the root curve in the second panel of Fig.~\ref{f:roots}, the left and right edge points of the curve are exceptional points because saddle-node bifurcations occur there (at a saddle-node bifurcation, the slope $dz_2/dz_1$ of the root curve is infinite). The root curve in the third panel of Fig.~\ref{f:roots} has four exceptional points. Two of them are in the lower and upper half planes where saddle-node bifurcations occur (i.e., where the slopes are infinite), while the other two are on the $z_1$ axis where pitchfork bifurcations occur. The root curve in the fourth panel of Fig.~\ref{f:roots} also has four exceptional points; three of them are where saddle-node bifurcations occur, while the fourth one is at the intersection between the upper and lower loops where a transcritical bifurcation occurs. The first panel of Fig.~\ref{f:roots} does not have exceptional points since no bifurcation occurs here.

One may notice that the first two root curves in Fig.~\ref{f:roots} resemble the shapes of rogue curves in Figs.~1 and 2. Indeed, the root curve of $\mathcal{P}_{\Lambda}^{[m]}(z_1, z_2)$ turns out to be closely related to certain rogue curves in DSI, as we will show in the next subsection.

\subsection{Asymptotic prediction of rogue curves under one large internal parameter}

In this subsection, we analytically predict the shapes of rogue waves in DSI. For this purpose, we make the following restrictions on parameters in DSI's rogue waves in Lemma 1.
\begin{enumerate}
\item  We set $p=1$ (the case of $p\ne 1$ will be discussed in a later section).
\item For a certain $m\ge 3$, $a_m$ is real, $a_m \gg 1$ when $m$ is even and $|a_m|\gg 1$ when $m$ is odd, and the other $a_j$ values in $\textbf{\emph{a}}$ are $O(1)$ and complex (the case of large negative $a_m$ when $m$ is even will be discussed in a later section).
\end{enumerate}

One may notice that the parameter choices (\ref{paraDSI1})-(\ref{paraDSI2}) for Figs.~1-2 meet these restrictions. In both cases, $p=1$. In addition, in (\ref{paraDSI1}), $\textbf{\emph{a}}=(0,0,0,5000)$, and $a_4=5000$ is large positive. In (\ref{paraDSI2}), $\textbf{\emph{a}}=(0,0,2000)$, and $a_3=2000$ is large.

Under the above parameter restrictions, we will show that rogue curves in DSI would appear, and their shapes in the $(x, y)$ plane would be predicted by the root curves of $\mathcal{P}_{\Lambda}^{[m]}(z_1, z_2)$. To present these results, we first introduce some definitions.

Let us define a curve $y=y_c(x)$ in the $(x, y)$ plane, which we call the critical curve, as
\[  \label{xyz1z2}
x=2z_1a_m^{2/m},\ \ \ y_c(x)=z_2 a_m^{1/m},
\]
where $(z_1, z_2)$ is every point on the root curve of $\mathcal{P}_{\Lambda}^{[m]}(z_1, z_2)$. Alternatively, the critical curve can be defined by the equation
\[ \label{xyz1z2bb}
\mathcal{P}_{\Lambda}^{[m]}\left(\frac{x}{2a_m^{2/m}}, \frac{y_c(x)}{a_m^{1/m}}\right)=0,
\]
or
\[ \label{ycxdef}
y_c(x)=a_m^{1/m}\mathcal{R}_{\Lambda, m}\left(\frac{x}{{2a_m^{2/m}}}\right),
\]
using the notation in Eq.~(\ref{Pmz1z20b}). This critical curve may also contain exceptional points where
\[
\frac{\partial}{\partial y_c}\mathcal{P}_{\Lambda}^{[m]}\left(\frac{x}{2a_m^{2/m}}, \frac{y_c}{a_m^{1/m}}\right)=0.
\]
Such points are also bifurcation points of the critical curve when this curve is viewed as a bifurcation diagram, because a dynamical system point of view similar to Eq.~(\ref{dz2dt}) also applies here. It is easy to see that an exceptional point $(x^{(e)}, y_c^{(e)})$ of the critical curve is related to an exceptional point $(z_1^{(e)}, z_2^{(e)})$ of the root curve as
\[
x^{(e)}=2a_m^{2/m}z_1^{(e)}, \quad y_c^{(e)}=a_m^{1/m}z_2^{(e)}.
\]
Thus, the two exceptional points are simply related by a stretching along the horizontal and vertical axes.

Under these definitions, we have the following theorem.

\begin{quote}
\textbf{Theorem 1.} Let $A_\Lambda(x,y,t)$ be a DSI's rogue wave with order-index vector $\Lambda=\left(n_1, n_2, \ldots, n_N  \right)$ in Eq.~(\ref{DSRWsolu1}) of Lemma~1. Under the parameter restrictions mentioned above and when time $t=\mathcal{O}(1)$, we have the following asymptotic result on the solution $A_\Lambda(x,y,t)$ in the $(x, y)$ plane for large $|a_m|$.
\begin{enumerate}
\item  If $(x, y)$ is not in the $O(1)$ neighborhood of the critical curve $y=y_c(x)$, then the solution $A_\Lambda(x,y,t)$ approaches the constant background $\sqrt{2}$ as $|a_m| \to +\infty$.
\item If $(x, y)$ is in the $O(1)$ neighborhood of the critical curve $y=y_c(x)$, but not in the $O(1)$ neighborhood of its exceptional points, then the solution $A_\Lambda(x,y,t)$ at large $|a_m|$ would asymptotically form a rogue curve $A_{R}(x, y, t)$, whose expression is
\begin{eqnarray}  \label{ARxyt}
A_{R}(x,y,t)=\sqrt{2} \left[1 + \frac{4{\rm{i}}t-1}{\left[y-y_c(x)\right]^2+4t^2+\frac{1}{4}}\right].
\end{eqnarray}
The error of this rogue curve approximation is $O(a_m^{-1/m})$. Expressed mathematically, when $(x, y_c(x))$ is not an exceptional point of the critical curve and $\left|y(x)-y_c(x) \right|=O(1)$, we have the following solution asymptotics
\[ \label{Theorem3asym}
A_{\Lambda}(x,y,t) = A_{R}(x,y,t) + O\left(|a_m|^{-1/m}\right), \quad |a_m| \gg 1.
\]
\end{enumerate}
\end{quote}

The proof of this theorem will be given in a later section.

Notice that $A_{R}(x, y, t)$ in Eq.~(\ref{ARxyt}) is the same as the Peregrine rogue wave of the nonlinear Schr\"odinger equation (along the $y$ direction), except for a $y$-directional shift. The peak location of $|A_{R}(x, y, t)|$ at each $y$ value is at $y=y_c(x)$. All these peak locations from different $y$ values fall precisely on the critical curve $y=y_c(x)$. Thus, we can say the critical curve $y=y_c(x)$ predicts the spatial location of the rogue curve. The full rogue curve surrounding that critical curve is predicted by the function $A_{R}(x, y, t)$. We will compare these predictions to the true solutions of Figs.~1-2 in the next section. The root curves of $\mathcal{P}_{\Lambda}^{[m]}(z_1, z_2)$ involved in Eq.~(\ref{xyz1z2}) for those predictions are precisely the ones shown in the left two panels of Fig.~\ref{f:roots}. In cases where the root curve is closed so that $z_1$ of the root curve is only on a limited interval (see the second panel of Fig.~\ref{f:roots} for an example), this $A_{R}(x, y, t)$ prediction would be only for a limited $x$ interval as well in view of Eq.~(\ref{xyz1z2}). Outside that $x$ interval, our prediction of $A_\Lambda(x, y, t)$ would be the background value $\sqrt{2}$ as long as $(x, y)$ is not in the $O(1)$ neighborhood of the critical curve $y=y_c(x)$, according to the first statement of Theorem~1.

The only $(x, y)$ places where Theorem 1 does not make a solution prediction are $O(1)$ neighborhoods of the exceptional points on the critical curve $y=y_c(x)$. In such special neighborhoods, a more elaborate analysis than the ones to be employed in this article is needed in order to predict the solution behavior there.

\section{Comparison between analytical predictions and true solutions}
In this section, we compare analytical predictions of rogue curves in Theorem 1 to true solutions.

First of all, parameter choices (\ref{paraDSI1})-(\ref{paraDSI2}) for Figs.~1-2 meet the assumptions of Theorem 1. Thus, we will compare Theorem 1's predictions on them to the true solutions shown in Figs.~1-2.

\begin{figure}[htb]
\begin{center}
\includegraphics[scale=0.5, bb=550 0 300 300]{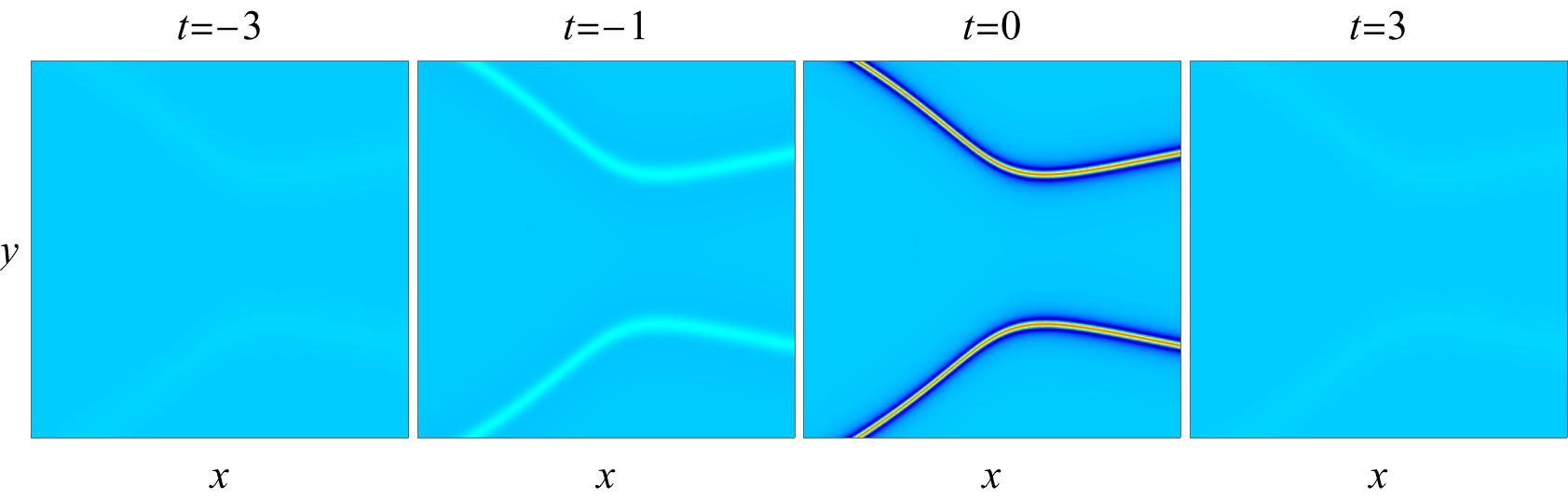}
\caption{Analytical predictions of the rogue curve for the parameter choices of (\ref{paraDSI1}) at four time values of $t=-3, -1, 0$ and 3. The $(x, y)$ intervals here are the same as those in Fig.~1 for easy comparison. } \label{f:roguecurvePred}
\end{center}
\end{figure}

For the first parameter choices (\ref{paraDSI1}),
\[
m=4, \quad a_m=5000, \quad \Lambda=(1, 4).
\]
In this case, the corresponding root curve of $\mathcal{P}_{\Lambda}^{[m]}(z_1, z_2)$ has been plotted in the first panel of Fig.~\ref{f:roots}. Using that root curve, we can obtain the predicted rogue curve $A_R(x, y, t)$ from Eqs.~(\ref{xyz1z2}) and (\ref{ARxyt}). At four time values of $t=-3, -1, 0$ and 3, corresponding to the time values chosen in Fig.~1, this $A_R(x, y, t)$ prediction is plotted in Fig.~\ref{f:roguecurvePred}. Comparing this figure to Fig.~1, we visually see that they closely match each other.

\begin{figure}[htb]
\begin{center}
\vspace{3cm}
\includegraphics[scale=0.6, bb=250 0 350 100]{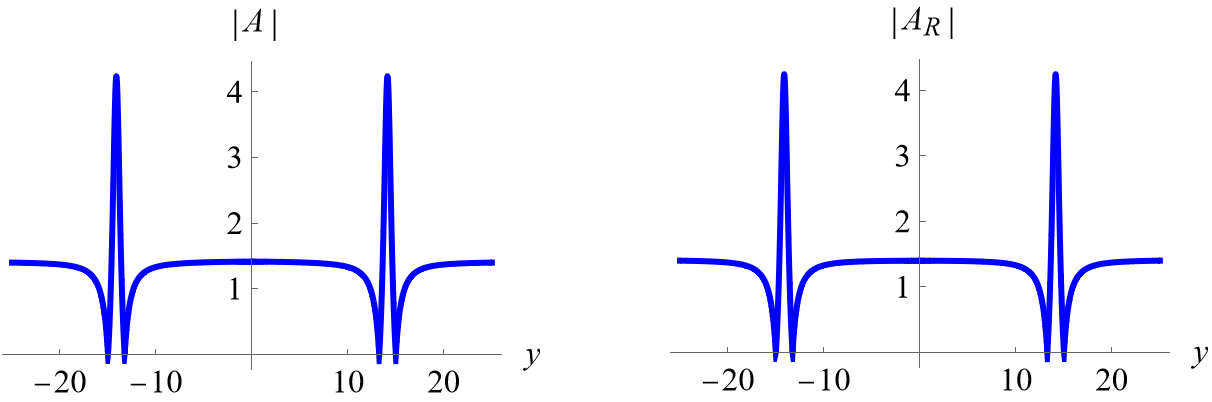}
\caption{The 1D comparison between the true rogue curve in Fig.~1 and its prediction at $x=t=0$.
Left panel: true solution $|A|$; right panel: predicted $|A_R|$. }  \label{f:1Dcomparison}
\end{center}
\end{figure}

To further compare the predicted and true solutions in this case, we set $x=t=0$, and compare the true and predicted $A(x, y, t)$ solutions versus the $y$ coordinate. This 1D comparison is shown in Fig.~\ref{f:1Dcomparison}. Again, this comparison shows very good agreement as well.

We have also verified the error decay rate of $O(a_m^{-1/m})$ in the neighborhood of the critical curve in Theorem~1 for the rogue curve solution in Fig.~1 by varying its large $a_4$ parameter and measuring the error between the prediction and the true solution. Details are omitted.

\begin{figure}[htb]
\begin{center}
\includegraphics[scale=0.5, bb=550 0 300 300]{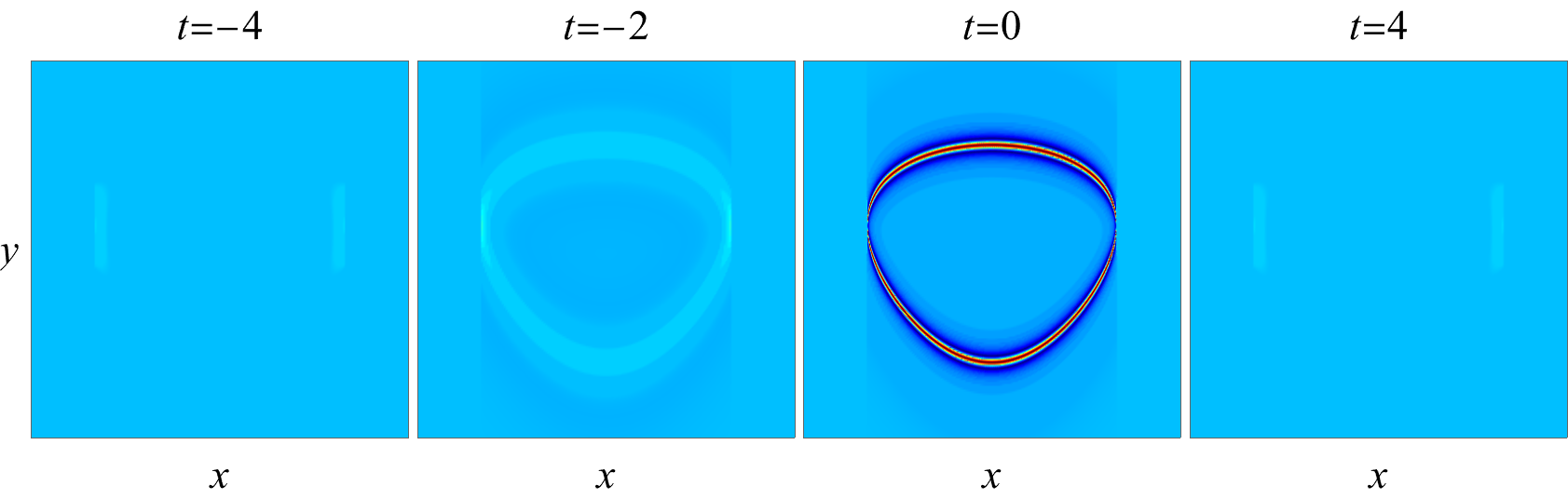}
\caption{Analytical predictions of the rogue ring for the parameter choices of (\ref{paraDSI2})
at four time values of $t=-4, -2, 0$ and 4. The $(x, y)$ intervals here are the same as those in Fig.~2 for easy comparison. } \label{f:rogueringPred}
\end{center}
\end{figure}

Next, we compare the true solution in Fig.~2 to our prediction for the second parameter choices (\ref{paraDSI2}). In this case,
\[
m=3, \quad a_m=2000, \quad \Lambda=(2, 3),
\]
and the corresponding root curve of $\mathcal{P}_{\Lambda}^{[m]}(z_1, z_2)$ has been plotted in the second panel of Fig.~\ref{f:roots}. Using that root curve, we obtain the predicted rogue curve $A_R(x, y, t)$ from Eqs.~(\ref{xyz1z2}) and (\ref{ARxyt}), which is a rogue ring. This $A_R(x, y, t)$ prediction only holds for the $x$ interval of $\left(2z_{1,L}a_m^{2/m}, 2z_{1,R}a_m^{2/m}\right)$, where $(z_{1,L}, z_{1,R})$ is the $z_1$ interval of the underlying root curve in the second panel of Fig.~\ref{f:roots}. For this root curve, $z_{1,R}=-z_{1,L}=3^{2/3}/2 \approx 1.0400$. Thus, the $x$ interval of this $A_R(x, y, t)$ prediction is $|x|< 6000^{2/3}\approx 330.19$. Outside this $x$ interval, we will use the uniform background $\sqrt{2}$ prediction for $A_\Lambda(x, y,t)$ according to the first statement of Theorem 1. At four time values of $t=-4, -2, 0$ and 4, corresponding to the time values chosen in Fig.~2, this rogue-ring $A_R(x, y, t)$ prediction is plotted in Fig.~\ref{f:rogueringPred}. Note that in this example, the critical curve $y=y_c(x)$ contains two exceptional points, which correspond to the left and right edge points of the rogue ring seen in the second and third panels of Fig.~\ref{f:rogueringPred}. According to Theorem 1, our predicted solutions in all four panels of Fig.~\ref{f:rogueringPred} are not expected to be valid in the $O(1)$ neighborhoods of those edge points.

Comparing our predicted solution in Fig.~\ref{f:rogueringPred} to the true one in Fig.~2, we visually see that the predicted rogue ring closely matches the true one in its shape and location. We have checked that the amplitudes on the predicted rogue ring closely match those on the true ring as well. For instance, at $x=t=0$, we have quantitatively compared the predicted and true $A(x, y, t)$ solutions versus $y$, similar to what we have done in Fig.~\ref{f:1Dcomparison} for the first example. We have found close agreement between prediction and the true solution in that 1D comparison (details of this 1D comparison are omitted for brevity).

The predicted solution in Fig.~\ref{f:rogueringPred} and the true one in Fig.~2 also have notable differences though, and those differences are mostly at or near the left and right edges of the rogue ring. At those edges, the true solution shows a lump there, which is very narrow and hardly visible at $t=0$ but becomes wider and more visible as $|t|$ increases. The predicted solution, however, does not exhibit such lumps. The reason for this difference is clearly due to the fact that those edge points are exceptional points of the critical curve, where our predicted solution does not hold according to Theorem 1. So, there are no contradictions between the analytical theory and the true solution here.

\begin{figure}[htb]
\begin{center}
\includegraphics[scale=0.5, bb=500 0 300 460]{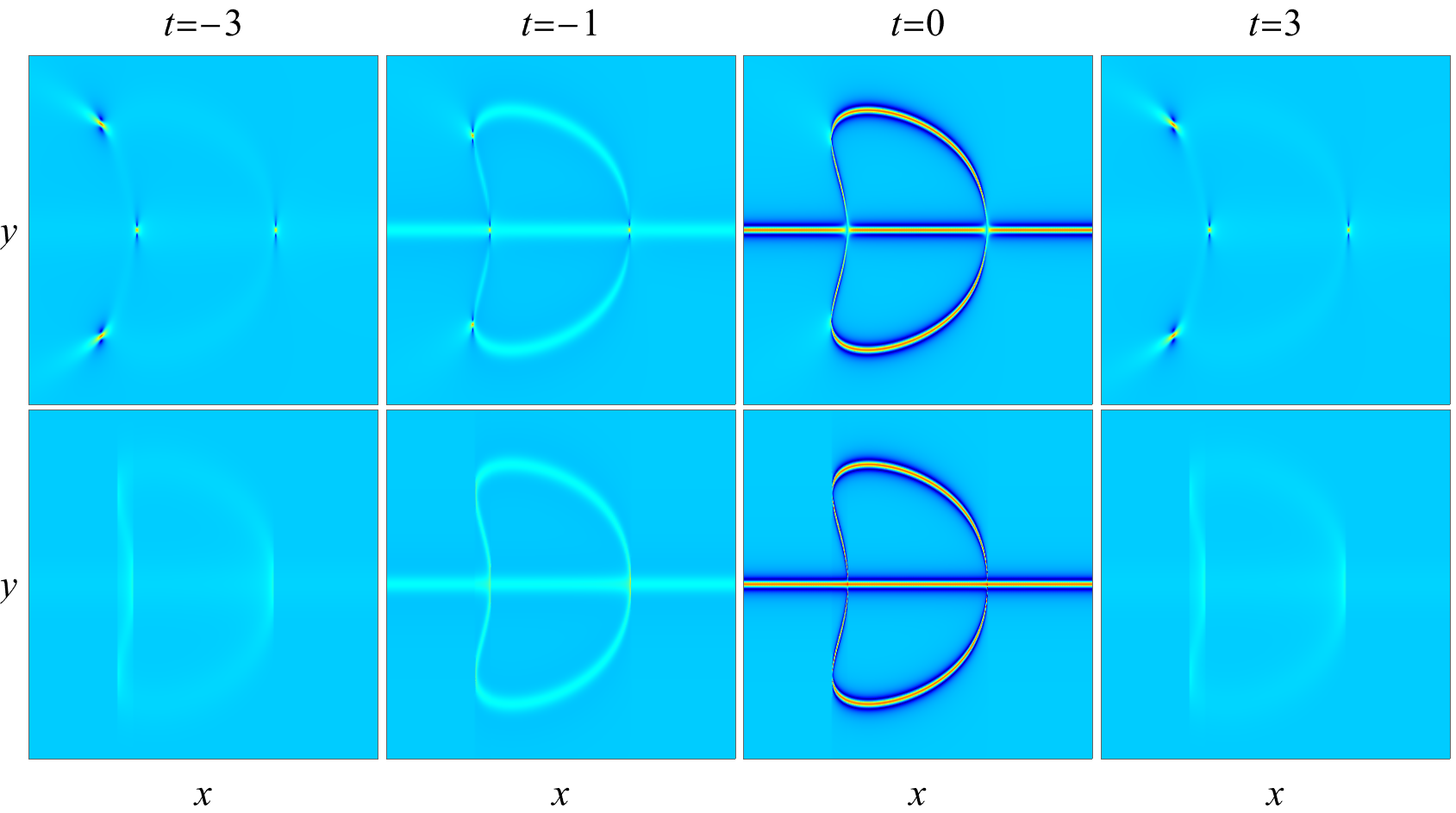}
\caption{Comparison on the rogue curve for the third example with parameter choices (\ref{paraDSI3}). Upper row: the true DSI solution $|A|$ at time values of $t=-3, -1, 0$ and 3. Lower row: theoretical predictions from Theorem 1 at the same time values. In all panels, $-500\le x \le 500$, and $-30\le y\le 30$.  } \label{f:example3}
\end{center}
\end{figure}

Next, we will do comparisons on two additional examples. The third example is where the parameter choices are
\[ \label{paraDSI3}
p=1, \quad  \Lambda=(2,4), \quad \textbf{\emph{a}}=(0,1,2{\rm{i}},5000).
\]
The corresponding true solution $|A|$ at four time values of $t=-3, -1, 0$ and 3 is plotted in the upper row of Fig.~\ref{f:example3}. It is seen that a rogue curve in the shape of a heart intersected by a line arises from a uniform background with four lumps on it. This rogue curve reaches peak amplitude at approximately $t=0$ and then fades away to that same background afterwards.

To compare this true solution to our prediction, we notice from Eq.~(\ref{paraDSI3}) that $m=4$ here since $a_4=5000$, which is the large parameter. The corresponding root curve for $\mathcal{P}_{\Lambda}^{[m]}(z_1, z_2)$, with $m=4$ and $\Lambda=(2, 4)$, has been plotted in the third panel of Fig.~\ref{f:roots}. Using that root curve and the $a_4$ value, we obtain the predicted rogue curve $A_R(x, y, t)$ from Eqs.~(\ref{xyz1z2}) and (\ref{ARxyt}), which is plotted in the lower row of Fig.~\ref{f:example3} at the same time values of the upper row. It is seen that the prediction matches the true solution pretty well. The main differences between them are at the four lumps, which are clearly at the exceptional points of the critical curve where the prediction in Theorem 1 does not hold. Thus such differences are not surprising.

\begin{figure}[htb]
\begin{center}
\includegraphics[scale=0.5, bb=400 0 500 300]{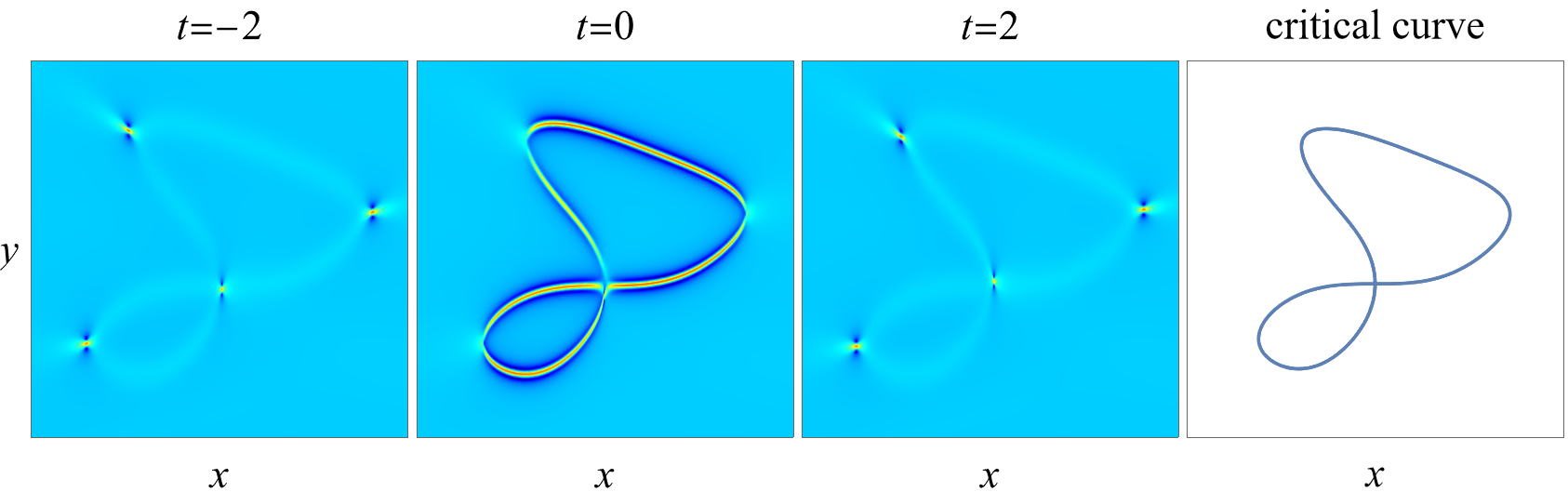}
\caption{Comparison on the rogue curve for the fourth example with parameter choices (\ref{paraDSI4}). Left three panels: the true DSI solution $|A|$ at three time values of $t=-2, 0$ and 2. Fourth panel: the underlying theoretical critical curve $y=y_c(x)$. In all panels, $-250\le x \le 250$, and $-25\le y\le 37$.  } \label{f:example4}
\end{center}
\end{figure}

The fourth example we examine is where the parameter choices are
\[ \label{paraDSI4}
p=1, \quad \Lambda=(4,5), \quad \textbf{\emph{a}}=(0,{\rm{i}},2{\rm{i}},3{\rm{i}},20000).
\]
The corresponding true solution $|A|$ at three time values of $t=-2, 0$ and 2 is plotted in the first three panels of Fig.~\ref{f:example4}. This time, a double rogue ring in a knot configuration arises from a uniform background with four lumps on it. This rogue curve reaches peak amplitude at approximately $t=0$ and then fades away to that same background afterwards.

To compare this true solution to our prediction, we see from Eq.~(\ref{paraDSI4}) that $m=5$ since $a_5=20000$ is large here. The corresponding root curve for $\mathcal{P}_{\Lambda}^{[m]}(z_1, z_2)$, with $m=5$ and $\Lambda=(4, 5)$, has been plotted in the last panel of Fig.~\ref{f:roots}. Using that root curve and the $a_5$ value, the critical curve $y=y_c(x)$ for this case can be obtained, which is shown in the last panel of Fig.~\ref{f:example4}. As we have mentioned earlier, the critical curve $y=y_c(x)$ predicts the spatial location of the rogue curve. When we compare this critical curve to the spatial location of the true rogue curve in the second panel of Fig.~\ref{f:example4}, they again match each other very well. The four lumps in the true solution are at the exceptional points of the critical curve, where Theorem 1 does not give an analytical prediction.

\section{Proof of analytical predictions in Theorem 1} \label{sec:proof}

In this section, we prove the analytical predictions in Theorem~1. Since $\epsilon=p=1$ and $a_1=0$, we see from Lemma 1 that
\[
x_1^\pm(k)=y\mp 2 \textrm{i} t \pm k, \quad x_2^+=\frac{1}{2}x +a_2, \quad x_2^-=(x_2^+)^*, \quad x_3^+=\frac{1}{6}y -\frac{4}{3} \textrm{i} t+a_3, \quad x_3^-=(x_3^+)^*,
\]
and so on. According to the assumptions of Theorem 1, for a certain integer $m\ge 3$, $a_m$ is real, $a_m \gg 1$ when $m$ is even and $|a_m|\gg 1$ when $m$ is odd, and the other $a_j$ values in $\textbf{\emph{a}}$ are $O(1)$ and complex. Suppose $x=O(a_m^{2/m})$, $y=O(a_m^{1/m})$, $t=O(1)$, and denote
\[ \label{xyz1z2b}
x = 2 z_1 a_m^{2/m}, \quad y = z_2 a_m^{1/m},
\]
where $z_1$ and $z_2$ are $O(1)$ and both real since $a_m^{1/m}$ is real due to the above assumptions on $a_m$. In this case,
\begin{eqnarray*}
&& S_n\left[\textbf{\emph{x}}^{+}(k)+\nu \textbf{\emph{s}}\right]=S_n ( y-2\textrm{i}t+k,\  \frac{1}{2}x+\nu s_2 +a_2, \cdots )\sim S_n(y, \frac{1}{2}x, 0, \cdots, 0, a_m, 0, \cdots) \\
&& = a_m^{n/m} S_n\left( y a_m^{-1/m},  \frac{1}{2}x a_m^{-2/m}, 0, \cdots, 0,  1,  0, \cdots  \right)
=a_m^{n/m} S_n\left(z_2, z_1, 0, \cdots, 0,  1,  0, \cdots  \right).
\end{eqnarray*}
Recall from the definition of Schur polynomials (\ref{Elemgenefunc}) that
\[
S_n\left(z_2, z_1, 0, \cdots, 0,  1,  0, \cdots  \right)=\mathcal{S}^{[m]}_n(z_1, z_2),
\]
where $\mathcal{S}^{[m]}_n(z_1, z_2)$ is as defined in Eq.~(\ref{AMthetak}). Thus,
\[ \label{SnplusRR}
S_n\left[\textbf{\emph{x}}^{+}(k)+\nu \textbf{\emph{s}}\right]\sim a_m^{n/m}\mathcal{S}^{[m]}_n(z_1, z_2), \quad |a_m|\gg 1.
\]
Similarly, we can also show that
\[ \label{SnminusRR}
S_n\left[\textbf{\emph{x}}^{-}(k)+\nu \textbf{\emph{s}}\right]\sim a_m^{n/m}\mathcal{S}^{[m]}_n(z_1, z_2), \quad |a_m|\gg 1.
\]

Now, we rewrite the determinant $\tau_k$ in Eq.~(\ref{deftaunk}) as a larger $(N+n_N+1)\times (N+n_N+1)$ determinant
\[ \label{3Nby3Ndet2}
\tau_k=\left|\begin{array}{cc}
\textbf{O}_{N\times N} & \Phi_{N\times (n_N+1)} \\
-\Psi_{(n_N+1)\times N} & \textbf{I}_{(n_N+1)\times (n_N+1)} \end{array}\right|,
\]
where
\[
\Phi_{i,j}=2^{-(j-1)} S_{n_i+1-j}\left(\textbf{\emph{x}}^{+}(k) + (j-1) \textbf{\emph{s}}\right), \quad \Psi_{i,j}=2^{-(i-1)} S_{n_j+1-i}\left(\textbf{\emph{x}}^{-}(k) + (i-1) \textbf{\emph{s}}\right),
\]
and vectors $\textbf{\emph{x}}^{\pm}(k)$ and $\textbf{\emph{s}}$ are as defined in Lemma 1. Using Laplace expansion, this larger determinant (\ref{3Nby3Ndet2}) can be rewritten as
\begin{eqnarray} \label{sigmanLap}
&& \hspace{-0.8cm} \tau_k=\sum_{0\leq\nu_{1} < \nu_{2} < \cdots < \nu_{N}\leq n_N}
\det_{1 \leq i, j\leq N} \left(\frac{1}{2^{\nu_j}} S_{n_i-\nu_j}({\textbf{\emph{x}}}^{+}(k)+\nu_j \textbf{\emph{s}}) \right)  \times \det_{1 \leq i, j\leq N}\left(\frac{1}{2^{\nu_j}}S_{n_i-\nu_j} ({\textbf{\emph{x}}}^{-}(k)+ \nu_j \textbf{\emph{s}} )\right).
\end{eqnarray}
The highest power term of $a_m$ in $\tau_k$ comes from the index choices of $\nu_{j}=j-1$. Then, using Eqs.~(\ref{SnplusRR})-(\ref{SnminusRR}), we can readily show that the highest $a_m$-power term of $\tau_k$ is
\begin{equation} \label{taukmax}
\tau_k \sim 2^{-N(N-1)} \hspace{0.05cm} a_m^{2\beta} \left[\mathcal{P}^{[m]}_{\Lambda}(z_1, z_2)\right]^2,
\quad \quad |a_m| \gg 1,
\end{equation}
where $\mathcal{P}^{[m]}_{\Lambda}(z_1, z_2)$ is the double-real-variable polynomial defined in Eq.~(\ref{DoubleRealPolydef}), and $\beta=(n_1+n_2+\dots+n_N-N(N-1)/2)/m$. Inserting this leading-order term of $\tau_k$ into Eq.~(\ref{DSRWsolu1}), we see that the solution $A_{\Lambda} (x,y,t)$ approaches $\sqrt{2}$ when $|a_m|\to \infty$, except at or near $(x, y)$ locations where
\[
\mathcal{P}^{[m]}_{\Lambda}(z_1, z_2)=0,
\]
or equivalently,
\[   \label{ycdef2}
\mathcal{P}_{\Lambda}^{[m]}\left(\frac{x}{2a_m^{2/m}}, \frac{y}{a_m^{1/m}}\right)=0
\]
in view of the connection (\ref{xyz1z2b}) between $(z_1, z_2)$ and $(x, y)$. The $(x, y)$ locations where Eq.~(\ref{ycdef2}) holds are the critical curve $y=y_c(x)$ as defined earlier in Eqs.~(\ref{xyz1z2bb})-(\ref{ycxdef}). Thus, if $(x, y)$ is not in the $O(1)$ neighborhood of this critical curve, the solution $A_{\Lambda} (x,y,t)$ approaches $\sqrt{2}$ for $|a_m|\gg 1$.

Next, we analyze the solution asymptotics in the $O(1)$ neighborhood of the critical curve. For this purpose, we denote $y=y_c+\hat{y}$, where $\hat{y}=O(1)$. Then, a more refined asymptotics for $S_n\left[\textbf{\emph{x}}^{+}(k)+\nu \textbf{\emph{s}}\right]$ is
\begin{eqnarray}
&& S_n\left[\textbf{\emph{x}}^{+}(k)+\nu \textbf{\emph{s}}\right]=S_n ( y-2\textrm{i}t+k,\  \frac{1}{2}x+\nu s_2 +a_2, \cdots )= S_n ( y_c+\hat{y}-2\textrm{i}t+k,\  \frac{1}{2}x, 0, \dots, 0, a_m, 0, \dots)\left[ 1+ \mathcal{O}( a_m^{-2/m}) \right] \nonumber \\
&& =a_m^{n/m} S_n\left[ y_c a_m^{-1/m} + (\hat{y}-2\textrm{i}t+k)a_m^{-1/m},  \frac{1}{2}x a_m^{-2/m}, 0, \cdots, 0,  1,  0, \cdots  \right]\left[ 1+ \mathcal{O}( a_m^{-2/m}) \right] \nonumber \\
&& =a_m^{n/m} S_n\left[ z_2 + (\hat{y}-2\textrm{i}t+k)a_m^{-1/m}, z_1, 0, \cdots, 0,  1,  0, \cdots  \right]\left[ 1+ \mathcal{O}( a_m^{-2/m}) \right].
\end{eqnarray}
Here, the point $(z_1, z_2)$ is on the root curve of $\mathcal{P}^{[m]}_{\Lambda}(z_1, z_2)=0$.

Now, we collect the dominant contributions in the Laplace expansion (\ref{sigmanLap}) for $\tau_k$ near the critical curve. There are two sources of contributions which are of the same order in $a_m$, one from the index choices of $\nu_j=j-1$, and the other from the index choices of $(\nu_1, \dots, \nu_N)=(0, 1, \dots, N-2, N)$. For the first index choice, using the above asymptotics of $S_n\left[\textbf{\emph{x}}^{+}(k)+\nu \textbf{\emph{s}}\right]$, we get
\begin{eqnarray*}
&& \left. \det_{1 \leq i, j\leq N} \left(\frac{1}{2^{\nu_j}} S_{n_i-\nu_j}({\textbf{\emph{x}}}^{+}(k)+\nu_j \textbf{\emph{s}}) \right)\right|_{\nu_j=j-1}=2^{-N(N-1)/2} \hspace{0.05cm} a_m^{\beta} \mathcal{P}^{[m]}_{\Lambda}\left(z_1, z_2+ (\hat{y}-2\textrm{i}t+k)a_m^{-1/m}\right)\left[ 1+  \mathcal{O}(a_m^{-2/m}) \right] \\
&& = 2^{-N(N-1)/2} \hspace{0.05cm} a_m^{\beta}  \left[\mathcal{P}^{[m]}_{\Lambda}\left(z_{1}, z_{2}\right) + \frac{\partial \mathcal{P}_{\Lambda}\left(z_{1}, z_{2}\right)}{\partial z_2}\left( \hat{y}-2\textrm{i}t+k \right)a_m^{-1/m} +\cdots \right] \left[ 1+ \mathcal{O}(a_m^{-2/m}) \right], \\
&& = 2^{-N(N-1)/2} \hspace{0.05cm} a_m^{\beta-1/m}\left[\frac{\partial \mathcal{P}^{[m]}_{\Lambda}\left(z_{1}, z_{2}\right)}{\partial z_2}\left( \hat{y}-2\textrm{i}t+k \right)
+\mathcal{O}( a_m^{-1/m})\right].
\end{eqnarray*}
Similarly, we get
\begin{eqnarray*}
&& \left. \det_{1 \leq i, j\leq N} \left(\frac{1}{2^{\nu_j}} S_{n_i-\nu_j}({\textbf{\emph{x}}}^{-}(k)+\nu_j \textbf{\emph{s}}) \right)\right|_{\nu_j=j-1}= 2^{-N(N-1)/2} \hspace{0.05cm} a_m^{\beta-1/m}\left[\frac{\partial \mathcal{P}^{[m]}_{\Lambda}\left(z_{1}, z_{2}\right)}{\partial z_2}\left( \hat{y}+2\textrm{i}t-k \right)
+\mathcal{O}( a_m^{-1/m})\right].
\end{eqnarray*}
Therefore, the contribution to the Laplace expansion (\ref{sigmanLap}) from the $\nu_j=j-1$ indices is
\[
\tau_k |_{\nu_j=j-1}=2^{-N(N-1)} \hspace{0.05cm} a_m^{2\beta-2/m}\left(  \frac{\partial \mathcal{P}^{[m]}_{\Lambda}}{\partial z_2}\right)^2
\left( \hat{y}-2\textrm{i}t+k \right)\left( \hat{y}+2\textrm{i}t-k \right)\left[ 1+ \mathcal{O}( a_m^{-1/m}) \right].
\]

For the other index choice of $(\nu_1, \dots, \nu_N)=(0, 1, \dots, N-2, N)$, using similar techniques and the relation (\ref{Smdz2}), we get
\[
\tau_k |_{\nu=(0, 1, \dots, N-2, N)}=2^{-N(N-1)-2} \hspace{0.05cm} a_m^{2\beta-2/m}\left(  \frac{\partial \mathcal{P}^{[m]}_{\Lambda}}{\partial z_2}\right)^2\left[ 1+ \mathcal{O}( a_m^{-1/m}) \right].
\]

Combining these two dominant contributions, we get
\[ \label{taukdom}
\tau_k=2^{-N(N-1)} \hspace{0.05cm} a_m^{2\beta-2/m} \left(  \frac{\partial \mathcal{P}^{[m]}_{\Lambda}}{\partial z_2} \right)^2 \left[ \left( \hat{y}-2\textrm{i}t+k \right)\left( \hat{y}+2\textrm{i}t-k \right)+ \frac{1}{4}\right]\left[ 1+ \mathcal{O}( a_m^{-1/m}) \right].
\]
Inserting this asymptotics of $\tau_k$ into Eq.~(\ref{DSRWsolu1}) and recalling $\hat{y}=y-y_c(x)$, we then get
\begin{eqnarray}
A(x,y,t)=\sqrt{2} \left[1 + \frac{4{\rm{i}}t-1}{\left[y-y_c(x)\right]^2+4t^2+\frac{1}{4}}\right]\left[ 1+ \mathcal{O}( a_m^{-1/m}) \right].
\end{eqnarray}
This is our asymptotic prediction of the $A(x, y, t)$ solution near the critical curve $y=y_c(x)$ as given in Theorem 1.

It is important to notice that the above predictions would be invalid near exceptional points of the critical curve, where $\partial \mathcal{P}^{[m]}_{\Lambda}/\partial z_2=0$. Near such points, the dominant contribution (\ref{taukdom}) to $\tau_k$ either vanishes or is of lower order in $a_m$, and a more advanced calculation would be needed to calculate $\tau_k$'s dominant contribution there.

This completes the proof of Theorem 1.

\section{Asymptotic predictions for $p\ne 1$} \label{sec7}

In Theorem 1, we set $p=1$. In this section, we discuss rogue curves when $p\ne 1$. Our assumptions on parameters $\textbf{\emph{a}}$ are the same as before, i.e., for a certain $m\ge 3$, $a_m$ is real, $a_m \gg 1$ when $m$ is even and $|a_m|\gg 1$ when $m$ is odd, and the other $a_j$ values in $\textbf{\emph{a}}$ are $O(1)$ and complex.

In this more general case,
\[
x_1^+(k)=c_{11} x +c_{12}y - c_{13} \textrm{i} t  + k, \quad x_2^+=c_{21}x +c_{22}y-c_{23} \textrm{i} t +a_2,
\]
and so on, where
\[
c_{11}=\frac{p-p^{-1}}{2}, \quad c_{12}=\frac{p+p^{-1}}{2}, \quad c_{13}=p^2+p^{-2},
\]
\[
c_{21}=\frac{p+p^{-1}}{4}, \quad c_{22}=\frac{p-p^{-1}}{4}, \quad c_{23}=p^2-p^{-2}.
\]
Then, for $|a_m|\gg 1$, $|x|\gg 1$, $|y|\gg 1$ and $t=O(1)$, we have
\begin{eqnarray*}
&& S_n\left[\textbf{\emph{x}}^{+}(k)+\nu \textbf{\emph{s}}\right]\sim S_n(c_{11} x +c_{12}y, c_{21}x +c_{22}y, 0, \cdots, 0, a_m, 0, \cdots).
\end{eqnarray*}
Setting
\[
c_{11} x +c_{12}y=z_2 a_m^{1/m}, \quad c_{21}x +c_{22}y=z_1a_m^{2/m},
\]
we get
\[
S_n\left[\textbf{\emph{x}}^{+}(k)+\nu \textbf{\emph{s}}\right]\sim S_n(z_2 a_m^{1/m}, z_1a_m^{2/m}, 0, \cdots, 0, a_m, 0, \cdots)=a_m^{n/m} S_n\left(z_2, z_1, 0, \cdots, 0,  1,  0, \cdots  \right)=a_m^{n/m}\mathcal{S}^{[m]}_n(z_1, z_2).
\]
Similarly,
\[
S_n\left[\textbf{\emph{x}}^{-}(k)+\nu \textbf{\emph{s}}\right]\sim a_m^{n/m}\mathcal{S}^{[m]}_n(z_1, z_2).
\]
Thus, using the Laplace expansion (\ref{sigmanLap}), we get
\begin{equation} \label{taukmax2}
\tau_k \sim 2^{-N(N-1)} \hspace{0.05cm} a_m^{2\beta} \left[\mathcal{P}^{[m]}_{\Lambda}(z_1, z_2)\right]^2,
\quad \quad |a_m| \gg 1,
\end{equation}
where $\beta$ has been defined below Eq.~(\ref{taukmax}). Inserting this leading-order term of $\tau_k$ into Eq.~(\ref{DSRWsolu1}), we see that the solution $A_{\Lambda} (x,y,t)$ approaches $\sqrt{2}$ when $|a_m|\to \infty$, except at or near $(x, y)$ locations where $\mathcal{P}^{[m]}_{\Lambda}(z_1, z_2)=0$, or equivalently,
\[   \label{ycdef2b}
\mathcal{P}_{\Lambda}^{[m]}\left(\frac{c_{21}x +c_{22}y}{a_m^{2/m}}, \frac{c_{11} x +c_{12}y}{a_m^{1/m}}\right)=0.
\]
This equation defines a critical curve where the rogue curve lies for $p\ne 1$. Using the notation of Eq.~(\ref{Pmz1z20b}), this critical curve can be written as
\[ \label{criticalcurvepn1}
c_{11} x +c_{12}y=\hat{y}_c(c_{21}x +c_{22}y),
\]
where
\[
\hat{y}_c(x)\equiv a_m^{1/m} \mathcal{R}_{\Lambda, m}\left(\frac{x}{a_m^{2/m}}\right).
\]
This critical curve is the counterpart of Eq.~(\ref{ycxdef}) for the $p=1$ case. In the $(x, y)$ plane, the shape of this critical curve is related to that of the previous one for $p=1$ through a linear transformation
\[ \label{lineartrans}
\left.\left(\begin{array}{c} x \\ y \end{array}\right)\right|_{p\ne 1}
=\left(\begin{array}{cc}2c_{21} & 2c_{22} \\ c_{11} & c_{12}\end{array}\right)^{-1}\left. \left(\begin{array}{c} x \\ y \end{array}\right)\right|_{p=1}.
\]

When $(x, y)$ is in the $O(1)$ neighborhood of this critical curve, we can further determine the asymptotic prediction for the solution, similar to what we have done for the $p=1$ case earlier. This predicted solution is found to be
\begin{eqnarray}  \label{ARxytb}
A_{R}(x,y,t)=\sqrt{2} \left[1 + \frac{2c_{13}{\rm{i}}t-1}{\left[c_{11}x+c_{12}y-\hat{y}_c(c_{21}x +c_{22}y)\right]^2+c_{13}^2t^2+\frac{1}{4}}\right]+O(|a_m|^{-1/m}).
\end{eqnarray}
When we compare the leading-order term of this prediction with the previous prediction (\ref{ARxyt}) for $p=1$ (at the same $\Lambda$ and $\textbf{\emph{a}}$ values), we see that the current prediction (for $p\ne 1$) spatially is the previous prediction of $p=1$ under a linear $(x, y)$ transformation (\ref{lineartrans}), plus a temporal rescaling. Thus, the current prediction can be viewed as a skewed Peregrine rogue wave (along the $c_{11}x+c_{12}y$ direction).

Similar to the $p=1$ case, our predictions would be invalid in the $O(1)$ neighborhood of exceptional points of the critical curve. These exceptional points are on the critical curve with the additional condition that
\[
\frac{\partial}{\partial z_2}\mathcal{P}_{\Lambda}^{[m]}\left(\frac{c_{21}x +c_{22}y}{a_m^{2/m}}, \frac{c_{11} x +c_{12}y}{a_m^{1/m}}\right)=0.
\]
It is easy to see that these exceptional points $(x^{(e)}, y_c^{(e)})$ of the critical curve are related to exceptional points $(z_1^{(e)}, z_2^{(e)})$ of the root curve as
\[
c_{21}x^{(e)} +c_{22}y^{(e)}=a_m^{2/m}z_1^{(e)}, \quad c_{11} x^{(e)} +c_{12}y^{(e)}=a_m^{1/m}z_2^{(e)}.
\]
It is important to note that in the present $p\ne 1$ case, an exceptional point of the critical curve may not be a bifurcation point of the critical curve when this critical curve is viewed as a bifurcation diagram in the $(x, y)$ plane, because the dynamical system point of view similar to Eq.~(\ref{dz2dt}) does not apply here. For example, a saddle-node bifurcation point (often an edge point) of the critical curve, in the $(x, y)$ plane, is generally not an exceptional point of the critical curve when $p\ne 1$.

\begin{figure}[htb]
\begin{center}
\includegraphics[scale=0.5, bb=500 0 300 450]{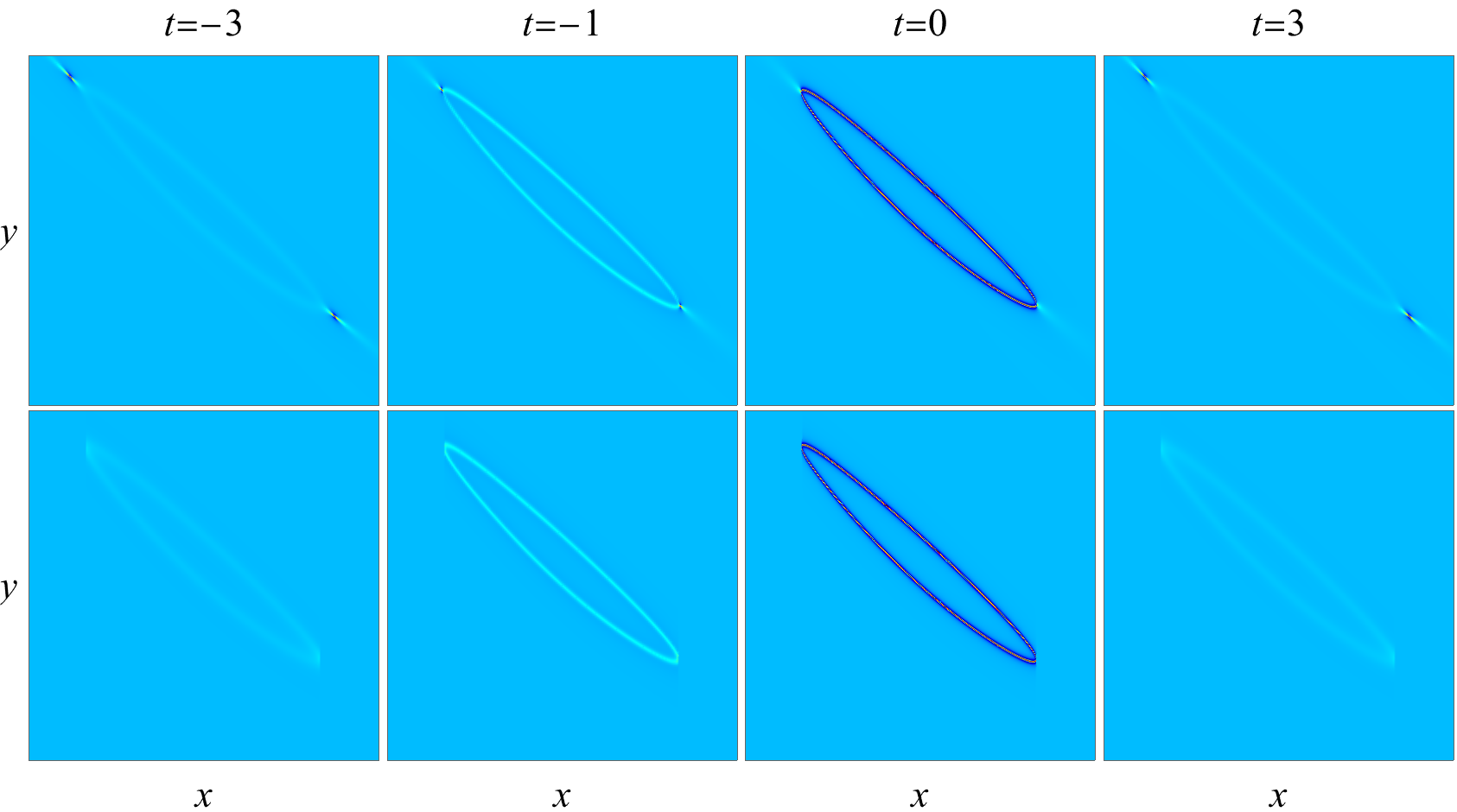}
\caption{Comparison on the rogue curve with $p\ne 1$ with parameter choices (\ref{paraDSI5}). Upper row: the true DSI solution $|A|$ at time values of $t=-3, -1, 0$ and 3. Lower row: theoretical predictions in Sec.~\ref{sec7} at the same time values. In all panels, $-500\le x \le 500$, and $-100\le y\le 100$.  } \label{f:example5}
\end{center}
\end{figure}

Now, we compare the above predictions with true solutions. For this purpose, we choose parameters in the DSI solution of Lemma~1 as
\[ \label{paraDSI5}
p=6/5, \quad \Lambda=(2, 3), \quad \textbf{\emph{a}}=(0,0,2000).
\]
The corresponding true solution $|A|$ at four time values of $t=-3, -1, 0$ and 3 is plotted in the upper row of Fig.~\ref{f:example5}. It is seen that a rogue curve in the shape of an elongated ring arises from a uniform background with two lumps on it. This rogue ring reaches peak amplitude at $t=0$ and then fades away to that same background afterwards.

Our predicted solution of $A(x, y, t)$, from Eq.~(\ref{ARxytb}) near the critical curve (\ref{criticalcurvepn1}) and $\sqrt{2}$ away from this curve, is plotted in the lower row of Fig.~\ref{f:example5} at the same time values of the true solution in its upper row. Comparing the predicted solution to the true solution, we see that the predicted rogue ring closely matches the true one in its shape and location. The main differences between the predicted and true solutions are at the two lumps, which are located at the exceptional points of the critical curve where our prediction does not hold.

\section{Extension to large negative $a_m$ when $m$ is even}

In Theorem~1, we assumed that when $m$ is even, $a_m$ would be large positive. It turns out that rogue curves can also arise in the case of large negative $a_m$ when $m$ is even. We will generalize our earlier results to this case in this section.

To study rogue curves for large negative $a_m$ with $m$ even, we will need to slightly modify the definition of double-real-variable polynomials $\mathcal{P}_{\Lambda}^{[m]}(z_1, z_2)$. In this case, we define
\begin{eqnarray} \label{DoubleRealPolydef2}
&& \hat{\mathcal{P}}_{\Lambda}^{[m]}(z_1, z_2) = \left| \begin{array}{cccc}
         \hat{\mathcal{S}}^{[m]}_{n_1}(z_1, z_2) & \hat{\mathcal{S}}^{[m]}_{n_1-1}(z_1, z_2) & \cdots &  \hat{\mathcal{S}}^{[m]}_{n_1-N+1}(z_1, z_2) \\
         \hat{\mathcal{S}}^{[m]}_{n_2}(z_1, z_2) & \hat{\mathcal{S}}^{[m]}_{n_2-1}(z_1, z_2) & \cdots &  \hat{\mathcal{S}}^{[m]}_{n_2-N+1}(z_1, z_2) \\
        \vdots& \vdots & \vdots & \vdots \\
         \hat{\mathcal{S}}^{[m]}_{n_N}(z_1, z_2) & \hat{\mathcal{S}}^{[m]}_{n_N-1}(z_1, z_2) & \cdots &  \hat{\mathcal{S}}^{[m]}_{n_N-N+1}(z_1, z_2)
       \end{array}
 \right|,
\end{eqnarray}
where Schur polynomials $\hat{\mathcal{S}}^{[m]}_k(z_1, z_2)$ are defined slightly differently from (\ref{AMthetak}) as
\begin{equation}\label{AMthetak2}
\sum_{k=0}^{\infty} \hat{\mathcal{S}}^{[m]}_k(z_1, z_2) \epsilon^k =\exp\left( z_2 \epsilon + z_1 \epsilon^{2} -   \epsilon^{m}  \right), \ \ \ m\geq3.
\end{equation}
For $p=1$, we define the critical curve $y=y_c(x)$ as
\[ \label{xyz1z2bb2}
\hat{\mathcal{P}}_{\Lambda}^{[m]}\left(\frac{x}{2|a_m|^{2/m}}, \frac{y_c(x)}{|a_m|^{1/m}}\right)=0.
\]
Then, the results of Theorem~1 (for $p=1$) would remain valid, i.e., a rogue curve would appear near this critical curve with its expression given by Eq.~(\ref{ARxyt}). For $p\ne 1$, the location and expression of the rogue curve could be similarly obtained from Sec.~\ref{sec7} by changing $\mathcal{P}_{\Lambda}^{[m]}$ to $\hat{\mathcal{P}}_{\Lambda}^{[m]}$ and $a_m$ to $|a_m|$ in Eq.~(\ref{ycdef2b}).

\section{Connections between rogue curves for different order-index vectors $\Lambda$}
In this section, we show that rogue curves for certain different order-index vectors $\Lambda=(n_1, n_2, \dots, n_N)$ are related to each other, using the theory of symmetric functions \cite{Murnaghan1937,Chakravarty2022KPI,Chakravarty2023KPI}.

For this purpose, it would be convenient to introduce the Young diagram $Y = (i_1, i_2, \dots, i_N)$, or a partition, of length $N$, which is a decomposition of a non-negative integer $M$ given by a sequence of descending non-zero numbers such that $i_1 \ge i_2 \ge \dots \ge i_N > 0$ and $|Y|:=i_1+\cdots+i_N=M$. The Schur function $W_Y(\emph{\textbf{x}})$, for vector $\emph{\textbf{x}}=(x_1, x_2, \dots)$ and Young diagram $Y = (i_1, i_2, \dots, i_N)$, is defined by
\[
W_Y(\emph{\textbf{x}})=\det_{1\le j, k\le N} [S_{i_j -j+k}(\emph{\textbf{x}})],
\]
where elementary Schur polynomials $S_j(\emph{\textbf{x}})$ are as defined in Eq.~(\ref{Elemgenefunc}).

The Young diagram $Y = (i_1, i_2, \dots, i_N)$ is often displayed as a rectangular array of left-justified boxes such that the $k$-th row from the top contains $i_k$ boxes, $k = 1,...,N$. Thus the Young diagram consists of $N$ rows and a total number of $M$ boxes. The conjugate $Y'$ of a partition $Y$ is a partition whose Young diagram is the transpose of the original one obtained by interchanging its rows and columns. Obviously, $|Y|=|Y'|=M,\ (Y')'=Y$. A partition $Y$ is called self-conjugate if $Y=Y'$.

A well known result from the theory of symmetric functions \cite{Murnaghan1937,Chakravarty2022KPI,Chakravarty2023KPI} is  the following \emph{involution symmetry} among Schur functions of a given partition $Y$ and its conjugate $Y'$:
\[\label{involution}
W_{Y'}(\emph{\textbf{x}})= W_Y(\omega(\emph{\textbf{x}})),\quad  \omega(x_j)= (-1)^{j-1}x_{j}.
\]
For example, when
\begin{equation*}
Y =(3, 2)=\ydiagram{3,2}\ ,
\end{equation*}
the conjugate partition $Y'$ is
\begin{equation*}
Y' =(2, 2, 1)=\ydiagram{2, 2, 1}\ .
\end{equation*}
The associated Schur functions are
\[
W_{Y}(\emph{\textbf{x}})=\frac{x_{1}^5}{24} + \frac{x_1^3 x_2}{6} + \frac{x_1 x_2^2}{2} - \frac{x_1^2 x_3}{2}  + x_2 x_3  - x_1 x_4,
\]
and
\[
W_{Y'}(\emph{\textbf{x}})=\frac{x_{1}^5}{24} - \frac{x_1^3 x_2}{6} + \frac{x_1 x_2^2}{2} - \frac{x_1^2 x_3}{2} - x_2 x_3 + x_1 x_4,
\]
which satisfy the involution symmetry (\ref{involution}).

Now, we relate the $\mathcal{P}_{\Lambda}^{[m]}(z_1, z_2)$ and $\hat{\mathcal{P}}_{\Lambda}^{[m]}(z_1, z_2)$ polynomials introduced earlier in Eqs.~(\ref{DoubleRealPolydef}) and (\ref{DoubleRealPolydef2}) to these Schur functions. It is easy to see that for $\Lambda=(n_1, n_2, \dots, n_N)$ with $n_1<n_2<\cdots <n_N$,
\[
\mathcal{P}_{\Lambda}^{[m]}(z_1, z_2)=W_{Y}(\emph{\textbf{x}}),
\]
where the Young diagram $Y = (i_1, i_2, \dots, i_N)$ is given by
\[\label{Indexrelation}
i_{N-(j-1)}=n_j-(j-1), \quad j=1, \dots, N,
\]
and $\emph{\textbf{x}}=(z_2, z_1, 0, \dots, 1, 0, \dots)$ with $1$ in its $m$-th element.
Similarly, $\hat{\mathcal{P}}_{\Lambda}^{[m]}(z_1, z_2)=W_{Y}(\hat{\emph{\textbf{x}}})$, where
$\hat{\emph{\textbf{x}}}=(z_2, z_1, 0, \dots, -1, 0, \dots)$ with $-1$ in its $m$-th element. We define the conjugate $\Lambda'$ of the order-index vector $\Lambda$ as one whose Young diagram is the conjugate of $\Lambda$'s Young diagram. For example, when $\Lambda=(4, 5)$, $Y=(4,4)$. Thus, $Y'=(2,2,2,2)$ and $\Lambda'=(2,3,4,5)$. $\Lambda$ is called self-conjugate if $\Lambda=\Lambda'$.

Due to the involution symmetry (\ref{involution}) of Schur functions, we find that
\[\label{symmetryodd}
\mathcal{P}_{\Lambda'}^{[m]}(z_1, z_2)= \mathcal{P}_{\Lambda}^{[m]}(-z_1, z_2), \quad \mbox{when $m$ is odd.}
\]
This means that root curves for the index vector $\Lambda$ and its conjugate vector $\Lambda'$ are related as a mirror reflection in $z_1$. Then, in view of the connection (\ref{xyz1z2}) between rogue curves of DSI and root curves of $\mathcal{P}_{\Lambda}^{[m]}(z_1, z_2)$, we conclude that when $p=1$, $m$ is odd and $a_m$ real and large in magnitude, the rogue curve $A_R(x, y, t)$ of DSI for the order-index vector $\Lambda$ would be a mirror reflection of the rogue curve in the $x$ variable for the conjugate order-index vector $\Lambda'$, i.e.,
\[ \label{RCsym1}
A_R(x, y, t)|_{\Lambda}=A_R(-x, y, t)|_{\Lambda'}, \quad \mbox{when $p=1$, $m$ is odd in large $|a_m|$.}
\]
For self-conjugate $\Lambda$ where $\Lambda'=\Lambda$, its rogue curve would be symmetric in $x$.

\begin{figure}[htb]
\begin{center}
\includegraphics[scale=0.4, bb=350 0 250 400]{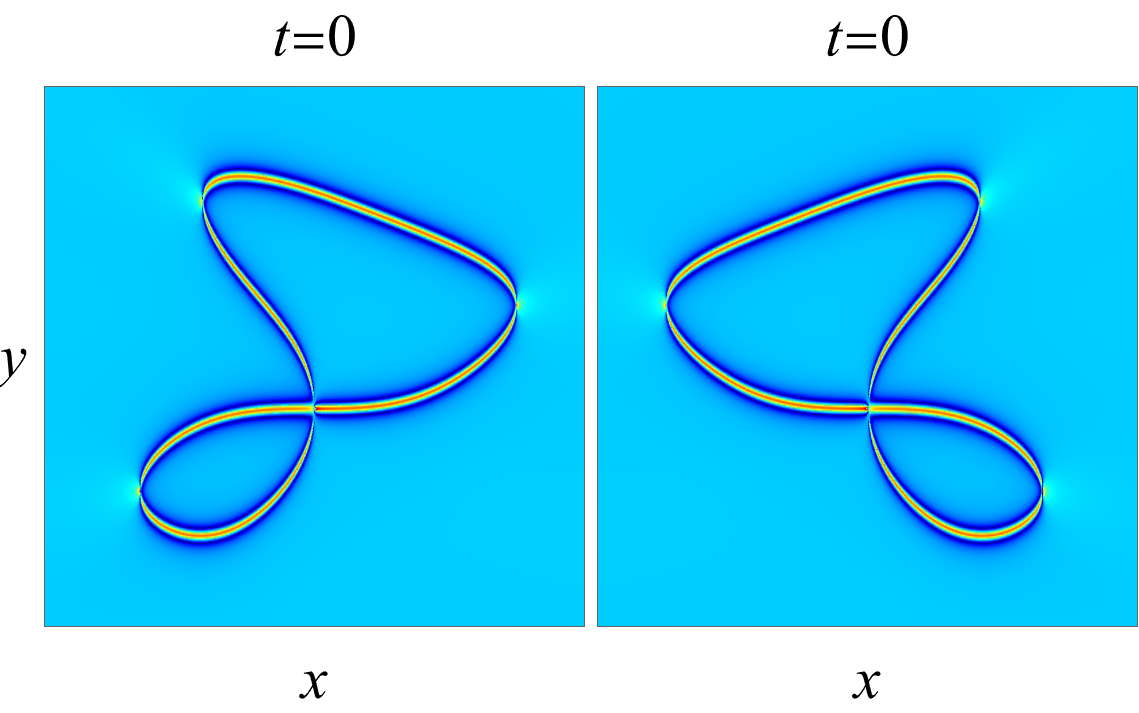}
\caption{Confirmation of symmetry (\ref{RCsym1}) between rogue curves $|A|$ for the order-index vector $\Lambda=(4, 5)$ (left) and its conjugate $\Lambda'=(2, 3, 4, 5)$ (right) with $p=1$, $\textbf{\emph{a}}=(0,0,0,0,20000)$, and $t=0$. These rogue curves are obtained from Lemma 1. In both panels, $-250\le x \le 250$, and $-25\le y\le 37$. } \label{f:RCsym1}
\end{center}
\end{figure}

To verify this connection (\ref{RCsym1}), we take $p=1$, and $\Lambda=(4,5)$. As we have mentioned earlier, for this $\Lambda$, its conjugate is $\Lambda'=(2,3,4,5)$. For these two order-index vectors $\Lambda$ and $\Lambda'$, we choose the same parameter vector $\textbf{\emph{a}}=(0,0,0,0,20000)$, where $a_5$ is real and large. Then, rogue curves $|A|$ for these $\Lambda$ and $\Lambda'$ at $t=0$ can be obtained from Lemma~1, which are displayed in Fig.~\ref{f:RCsym1}. One can see that these curves are indeed a mirror reflection of each other in $x$, confirming the above symmetry (\ref{RCsym1}).

When $m$ is even in the large real $a_m$ parameter (in magnitude), rogue curves for $\Lambda$ and $\Lambda'$ are also related, but the $a_m$ parameter for $\Lambda$ and $\Lambda'$ should have opposite signs, i.e., the symmetry now becomes
\[ \label{RCsym2}
A_R(x, y, t)|_{\Lambda, a_m}=A_R(-x, y, t)|_{\Lambda', -a_m}, \quad \mbox{when $p=1$, $m$ is even in large $|a_m|$.}
\]
We note that except for $a_m$, the other $O(1)$ parameters in the vectors $\textbf{\emph{a}}=(0, a_2, \dots, a_{n_N})$ for these two rogue waves of $\Lambda$ and $\Lambda'$ do not need to be opposite of each other and can be totally independent of each other. The connection (\ref{RCsym2}) can be seen from the symmetry
\[\label{symmetryeven}
\mathcal{P}_{\Lambda'}^{[m]}(z_1, z_2)=\hat{\mathcal{P}}_{\Lambda}^{[m]}(-z_1, z_2), \quad \mbox{when $m$ is even,}
\]
which comes from the involution symmetry (\ref{involution}) when $m$ is even. Here, $\hat{\mathcal{P}}_{\Lambda}^{[m]}$ is as defined in Eq.~(\ref{DoubleRealPolydef2}).

\begin{figure}[htb]
\begin{center}
\includegraphics[scale=0.4, bb=350 0 250 400]{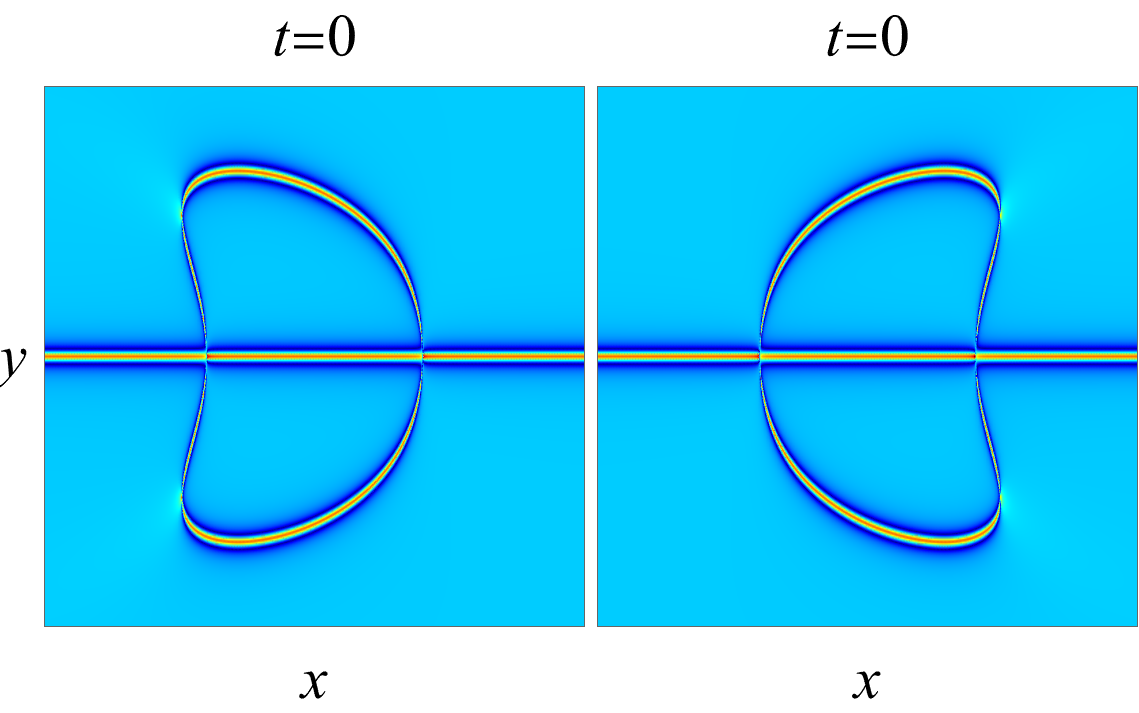}
\caption{Confirmation of symmetry (\ref{RCsym2}) between rogue curves $|A|$ for the order-index vector $\Lambda=(2, 4)$, $\textbf{\emph{a}}=(0,0,0,5000)$ (left), and its conjugate $\Lambda'=(1,3,4)$, $\textbf{\emph{a}}=(0,0,0,-5000)$ (right) with $p=1$ and $t=0$. These rogue curves are obtained from Lemma 1. In both panels, $-500\le x \le 500$, and $-30\le y\le 30$.} \label{f:RCsym2}
\end{center}
\end{figure}

To verify this rogue curve connection (\ref{RCsym2}), we take $p=1$, and $\Lambda=(2,4)$. For this $\Lambda$, its conjugate is $\Lambda'=(1,3,4)$. For these two order-index vectors $\Lambda$ and $\Lambda'$, we choose $\textbf{\emph{a}}=(0,0,0,5000)$ for $\Lambda$ and $(0,0,0,-5000)$ for $\Lambda'$, whose $a_4$ parameters are real and opposite of each other. Then, rogue curves $|A|$ for these two sets of parameters at $t=0$ can be obtained from Lemma~1, which are displayed in Fig.~\ref{f:RCsym2}. One can see that these curves are indeed a mirror reflection of each other in $x$, confirming the above symmetry (\ref{RCsym2}).

When $p\ne 1$, rogue curves for $\Lambda$ and $\Lambda'$ would be related through the symmetry (\ref{RCsym1}) or (\ref{RCsym2}) under a linear transformation, which can be seen from Eq.~(\ref{lineartrans}).

\section{Conclusions and discussions}
In this article, we have reported new rogue wave patterns whose wave crests form closed or open curves in the spatial plane in the Davey-Stewartson I equation. The shapes of these rogue curves, such as rings, double rings and others, are striking, and their appearance in the Davey-Stewartson I equation is a significant phenomenon. Analytically, we reveal that such rogue curves would appear when an internal parameter in bilinear expressions of the rogue waves are real and large. Performing large-parameter asymptotic analysis, we have discovered that these rogue curves can be predicted by root curves of certain types of double-real-variable polynomials. We have also compared our analytical predictions of rogue curves to true solutions and demonstrated good agreement between them.

An interesting question is whether such rogue curves would also appear in other (2+1)-dimensional integrable systems, such as the (2+1)-dimensional three-wave resonant interaction system. This question will be left to future studies.

\section*{Acknowledgment}
The work of B.Y. was supported in part by the National Natural Science Foundation of China (GrantNo.12201326), and the work of J.Y. was supported in part by the National Science Foundation (U.S.) under award number DMS-1910282.

\section*{Appendix}
General rational solutions (including rogue waves) in the Davey-Stewartson I eqaution (\ref{DS}) were derived in \cite{OhtaYangDSI}, but those solutions involved differential operators and were not explicit. In this appendix, we present explicit expressions of these general rational solutions and their brief proof.

\begin{quote}
\textbf{Lemma 2} The Davey-Stewartson I eqaution (\ref{DS}) admits general rational solutions
\begin{eqnarray}
  && A_{\Lambda}(x,y,t)= \sqrt{2}\frac{g}{f}, \label{DSRWsolu1b} \\
  && Q_{\Lambda}(x,y,t)= 1-2 \epsilon \left( \log f \right)_{xx}, \label{DSRWsolu2b}
\end{eqnarray}
where $N$ is a positive integer, $\Lambda=(n_1, n_2, \dots, n_N)$ is an order-index vector, each $n_i$ is a nonnegative integer, $n_1<n_2<\cdots <n_N$,
\[ \label{diffopesolufN2}
f=\tau_{0}, \quad g=\tau_{1},
\]
\[ \label{deftaunk2b}
\tau_{k}=
\det_{
\begin{subarray}{l}
1\leq i, j \leq N
\end{subarray}
}
\left(
\begin{array}{c}
m_{i,j}^{(k)}
\end{array}
\right),
\]
the matrix elements $m_{i,j}^{(k)}$ of $\tau_{k}$ are defined by
\[ \label{Schmatrimnijb}
m_{i,j}^{(k)}=\sum_{\nu=0}^{\min(n_{i}, n_{j})}\left(\frac{1}{p_{i}+p_{j}^*}\right) \left[ \frac{ p_{i} p_{j}^* }{(p_{i}+p_{j}^*)^2}  \right]^{\nu} \hspace{0.06cm} S_{n_{i}-\nu}[\textbf{\emph{x}}^{+}_{i,j}(k) +\nu \textbf{\emph{s}}_{i,j} + \textbf{\emph{a}}_i] \hspace{0.06cm} S_{n_{j}-\nu}[\textbf{\emph{x}}^{-}_{j,i}(k) + \nu \textbf{\emph{s}}^*_{j,i}+ \textbf{\emph{a}}_j^*],
\]
vectors $\textbf{\emph{x}}^{\pm}_{i,j}(k)=\left( x_{1,i,j}^{\pm}, x_{2,i,j}^{\pm},\cdots \right)$ are
\begin{eqnarray}
&&x_{r,i,j}^{+}(k)= \frac{(-1)^r}{r!p_i} x_{-1} + \frac{(-2)^r}{r!p_i^2} x_{-2} + \frac{1}{r!} p_{i} x_1 + \frac{2^r}{r!} p_{i}^2 x_2  + k \delta_{r, 1} - c_{r, i,j},    \label{xrijplusb} \\
&&x_{r,i,j}^{-}(k)= \frac{(-1)^r}{r!p_i^*} x_{-1} + \frac{(-2)^r}{r!(p_i^*)^2} x_{2} + \frac{1}{r!} p_{i}^* x_1 + \frac{2^r}{r!} (p_{i}^*)^{2} x_{-2}  - k \delta_{r, 1} - c_{r, i,j}^*,
\end{eqnarray}
\begin{equation} \label{x1x2xyt2}
\begin{array}{ll}
x_1=\frac{1}{2}(x+y), & x_{-1}=\frac{1}{2}\epsilon (x-y), \\ [5pt]
x_{2}=-\frac{1}{2}\textrm{i}t, & x_{-2}=\frac{1}{2}\textrm{i}t,
\end{array}
\end{equation}
$p_i$ are free complex constants, $\delta_{r, 1}$ is the Kronecker delta function, $\textbf{\emph{s}}_{i,j}=(s_{1,i,j}, s_{2,i,j}, \cdots)$, $c_{r,i,j}$ and $s_{r,i,j}$ are coefficients from the expansions
\[ \label{csexpansion}
\ln \left[ \frac{p_i  e^{\kappa} +p_j^*}{p_i+p_j^*} \right]= \sum_{r=1}^{\infty}c_{r,i,j} \kappa^r, \ \ \ \ln \left[\frac{p_i + p_j^* }{\kappa} \left( \frac{ e^\kappa -1}{p_i e^\kappa + p_j^*} \right)  \right] = \sum_{r=1}^{\infty}s_{r,i,j} \kappa^r,
\]
vectors $\textbf{\emph{a}}_i$ are
\[ \label{aivector}
\textbf{\emph{a}}_i=\left(a_{i,1}, a_{i,2}, \ldots, a_{i,n_i} \right),
\]
and $a_{i,j} \hspace{0.05cm} (1 \leq i \leq N, 1 \leq j \leq n_i)$ are free complex constants.
\end{quote}

\textbf{Note.} In the above lemma, rogue waves would be obtained when all $p_i$ are real. If $p_i$ are not real, the rational solutions (\ref{DSRWsolu1})-(\ref{DSRWsolu2}) would be soliton or multi-solitons on a constant background, not rogue waves \cite{OhtaYangDSI}.

\vspace{0.1cm}\noindent
\textbf{Proof.} From the appendix of Ref.~\cite{OhtaYangDSI}, we know that DSI admits the following rational solutions in differential operator form,
\[
A_{\Lambda}(x,y,t)= \sqrt{2}\frac{g}{f}, \quad Q_{\Lambda}(x,y,t)= 1-2 \epsilon \left( \log f \right)_{xx},
\]
where $\Lambda=(n_1, n_2, \dots, n_N)$, $N$ is the length of $\Lambda$, each $n_i$ is a nonnegative integer, $n_1<n_2<\cdots <n_N$,
\[
f=\tau_{0}, \quad g=\tau_{1},
\]
\[
\tau_{k}=
\det_{
\begin{subarray}{l}
1\leq i, j \leq N
\end{subarray}
}
\left(
\begin{array}{c}
m_{i,j}^{(k)}
\end{array}
\right),
\]
the matrix elements $m_{i,j}^{(k)}$ of $\tau_{k}$ are defined by
\[ \label{mijdiff11}
m_{i,j}^{(k)}=
  \frac{\left(p\partial_{p}\right)^{n_{i}}}{ (n_{i}) !}\frac{\left(q \partial_{q}\right)^{n_{j}}}{(n_{j}) !} \left.
  \left[ \frac{1}{p + q}\left(-\frac{p}{q}\right)^{k} e^{\Theta_{i,j}(x,y,t)}\right]\ \right|_{p=p_{i}, \ q=q_j},
\]
\[
\Theta_{i,j}(x,y,t)=  \left( \frac{1}{p^2} - \frac{1}{q^2} \right)x_{-2}+  \left( \frac{1}{p} + \frac{1}{q} \right)x_{-1} + (p+q) x_1+ (p^2-q^2)x_2 + \sum _{r=1}^\infty  a_{r,i} \ln^r \left[ \frac{p}{p_{i}} \right] + a^*_{r,j}  \ln^r \left[ \frac{q}{q_{j}} \right], \nonumber
\]
variables $(x_{-1}, x_{-2}, x_1, x_2)$ are related to $(x, y, t)$ by Eq.~(\ref{x1x2xyt2}), $p_j$ are free complex constants, $q_j=p_{j}^*$, and $a_{r,i} \hspace{0.05cm} (r=1, 2, \dots)$ are free complex constants. The main difference between this solution form and that in Ref.~\cite{OhtaYangDSI} is a simpler parameterization, which leads to simpler solution expressions.

The next step is to remove the differential operators in Eq.~(\ref{mijdiff11}) and derive explicit expressions for the matrix element $m_{i,j}^{(k)}$. The procedure to do this is very similar to that we used in Refs.~\cite{OhtaJY2012,YangYang3wave}. Performing such calculations, we can show that $m_{i,j}^{(k)}$ is as given in Eq.~(\ref{Schmatrimnijb}) of Lemma 2. This completes the brief proof of Lemma 2.

Lemma 1 in the main text is a special case of Lemma 2. To get Lemma 1, we set all $p_i$ to be the same and real in Lemma 2 and denote $p_i=p$, where $p$ is a real parameter. In addition, we require $a_{i,j}$ in Eq.~(\ref{aivector}) to be independent of the $i$ index. In this case, since the length of vector $\textbf{\emph{a}}_{i}$ is $n_i$, and $n_1<n_2<\cdots<n_N$, then, each $\textbf{\emph{a}}_{i}$ for $i<N$ is just a truncation of the longest vector $\textbf{\emph{a}}_{\Lambda}$. Since every $\textbf{\emph{a}}_{i}$ can be extended to the full $\textbf{\emph{a}}_{\Lambda}$, and the extended parts are dummy parameters which do not appear in the actual solution formulae, by performing this $\textbf{\emph{a}}_{i}$ extension, we can say all $\{\textbf{\emph{a}}_{i}\}$ vectors are the same in this case and thus denote
\[
\textbf{\emph{a}}_{i}=\textbf{\emph{a}}=(a_1, a_2, a_3, \dots, a_{n_N}).
\]

Under the above parameter restrictions, $c_{r,i,j}$ and $s_{r,i,j}$ in Lemma 2 are independent of the $(i, j)$ indices, i.e., $c_{r,i,j}=c_r$ and $s_{r,i,j}=s_r$. Similarly, $\textbf{\emph{x}}^{\pm}_{i,j}$ are independent of the $(i, j)$ indices too.
In addition, the $s_j$ expansion in Eq.~(\ref{csexpansion}) reduces to (\ref{schurcoeffsr}) in Lemma 1. We further lump the
$\textbf{\emph{a}}$ vector in Eq.~(\ref{Schmatrimnijb}) into $\textbf{\emph{x}}^{+}$ and the $\textbf{\emph{a}}^*$ vector into $\textbf{\emph{x}}^{-}$. In addition, we lump the $c_r$ parameter in Eq.~(\ref{xrijplusb}) into $a_r$ and
the $c_r^*$ parameter into $a_r^*$. After these treatments, we obtain Lemma 1 from Lemma 2.

\end{document}